\documentclass[a4paper,twoside,11pt]{article}
\usepackage[utf8]{inputenc}

\usepackage{amsmath}
\usepackage{mathptmx}
\DeclareMathAlphabet{\mathcal}{OMS}{cmsy}{m}{n}

\usepackage{graphicx}
\usepackage{subeqnarray}
\usepackage{xcolor}
\DeclareMathOperator*{\argmax}{\arg\max}   
\usepackage[final]{pdfpages}
\usepackage{graphicx}
\usepackage{dsfont}
\usepackage{subfig}
\usepackage[font=small,labelfont=bf]{caption}

\usepackage{geometry}
\usepackage{graphicx}
\usepackage[sort&compress,numbers]{natbib}
\usepackage{enumerate}
\usepackage{wasysym}
\usepackage[textsize=footnotesize]{todonotes}
\usepackage{url}

\usepackage{bm}

\definecolor{darkblue}{rgb}{0,0,0.6}
\definecolor{darkred}{rgb}{0.6,0,0}

\usepackage[colorlinks=true, urlcolor=black, citecolor=black, linkcolor=black, hyperfootnotes=false]{hyperref}

\usepackage{authblk}

\usepackage[bitstream-charter]{mathdesign}

\usepackage{natbib}
\usepackage{graphicx}

\newcommand{\IT}{\textrm{IT}}
\newcommand{\NT}{\textrm{NT}}
\newcommand{\MM}{\textrm{MM}}

\newcommand{\avg}{\mathbb{E}}
\newcommand{\filt}{\mathcal{I}}

\newcommand{\vecone}{\bm{1}}
\newcommand{\vecdmd}{\mathbf{R}}

\newcommand{\matprop}{\mathsf{G}}
\newcommand{\matsymprop}{\mathsf{G}^{\textrm{sym}}}
\newcommand{\matfcast}{\mathsf{F}}

\newcommand{\matL}{\mathsf{L}}

\newcommand{\matdmd}{\mathsf{R}}
\newcommand{\id}{\mathsf{I}}

\newcommand{\fut}{{/t}}
\newcommand{\past}{{t}}

\newcommand{\autocvol}{\Omega}

\newcommand{\util}{U}

\DeclareMathOperator{\1}{\mathsf{U}}
\graphicspath{{figures/}} 

\usepackage{authblk}

\title{Agent based propagators}

\author[1,2]{Michele Vodret\footnote{\url{mvodret@gmail.com}}}
\author[3]{Iacopo Mastromatteo}
\author[3]{Bence T\'oth}
\author[1,2,3]{Michael Benzaquen}
\affil[1]{Chair of Econophysics \& Complex Systems, Ecole polytechnique, 91128 Palaiseau Cedex, France}
\affil[2]{Ladhyx, UMR CNRS 7646, Ecole polytechnique, 91128 Palaiseau Cedex, France}
\affil[3]{Capital Fund Management, 23-25, Rue de l’Université 75007 Paris, France}
\date{\today \vspace{-1cm}}
\title{A Stationary Kyle Setup:
Microfounding propagator models}
\date{\today}

\begin{document}

\maketitle

\begin{abstract}

We provide an economically sound micro-foundation to linear price impact models, by deriving them as the  equilibrium of a suitable agent-based system. In particular, we retrieve the so-called 
\emph{propagator model} as the high-frequency limit of a generalized Kyle model, in which the assumption of a terminal time at which fundamental information is revealed is dropped. This allows to describe a stationary market populated by asymmetrically-informed rational agents. We investigate the stationary equilibrium of the model, and show that the setup is compatible with universal price diffusion at small times, and non-universal mean-reversion at time scales at which fluctuations in fundamentals decay.
Our model suggests that at high frequency one should observe a quasi-permanent impact component, driven by slow fluctuations of fundamentals, and a faster transient one, whose timescale should be set by the persistence of the order flow.

\end{abstract}
\providecommand{\keywords}[1]{\textbf{Keywords:} #1}

\keywords{Market microstructure, price impact, statistical inference.}

\tableofcontents

\setlength{\parskip}{\medskipamount}

\section{Introduction}
Financial markets are designed to achieve two seemingly unrelated goals: they allow market participants to find other agents with whom to transact (thereby solving a \emph{liquidity} problem), and at the same time they allow to discover the price at which such transactions should take place (thereby solving an \emph{information}-related task).

The Efficient Market Hypothesis (EMH) states that prices integrate all information that is publicly available~\cite{malkiel2003efficient}. 
If this is the case, there can be no forecastable structure in asset returns for agents in possession  of public information only. 
Historically, the EMH was first rationalized theoretically with the introduction of the Rational Expectation Hypothesis (REH). According to the REH all agents are rational and perfectly informed about the other players' strategies. This hypothesis is appealing since it allows to build analytically tractable setups \cite{O'Harabook} in which financial markets are able to deliver the promise they were conceived for, once some exogenous source of dynamics is injected into the system, thus preventing no-trade theorems. It has also important drawbacks: for example, the REH implies that the value of a risky asset is completely determined by its fundamental price, equal to the present discounted value of the expected stream of future dividends. As already argued by Shiller \cite{Shiller}, the excess volatility puzzle, i.e., the fact that the price deviates substantially from the fundamental value, cannot be explained by the REH.  Nevertheless, the REH is still considered the main expectation formation paradigm in many economic circles ~\cite{Econ_revolution}.

 An important class of REH models is the so-called Information-Based Models.  These models typically involve the presence of agents that trade due to exogenous reasons (\emph{noise traders}) that use financial markets in order to find counterparties for satisfying needs that come from \emph{outside} of the market, and arbitrageurs that possess privileged information on the traded goods (\emph{informed traders}), and thus choose to transact whenever they expect to use their informational advantage in their favor.
From this perspective, informed traders provide a service (making prices informative) that noise traders can choose to pay in order to be granted access to liquidity. 
To lubricate this mechanism, dealers (\emph{market makers}) are typically required: instead of letting noise traders and informed traders interact directly, market makers can temporarily incorporate the imbalance in the trading pressure, accepting to bear inventory risk for a limited time under the promise of some reward (bid-ask spread, rebate fees). Their activity allows to defer in time the moment at which the initial buyer and the final seller meet, thus enabling both informed  and noise traders to find more easily possible counter-parties.


A particularly successful class of models to describe statistical regularities in financial markets involves the notion of \emph{propagator}, a linear kernel used in autoregressive models that couples price changes to past order flow imbalances. In this setting, the (discounted) price of a good at time $t$, which we denote  $p_t$, can be expressed as a function of the past signed order flow imbalance $q_t$ as:
\begin{equation}
    \label{eq:prop}
    p_t := \sum_{t' = -\infty}^t G_{t-t'} q_{t'} \, ,
\end{equation}
where the causal kernel $G_{t}$ is the propagator.\footnote{Note that we are omitting from Eq.~\eqref{eq:prop} a residual noise term, that can be easily restored in order to account for price changes that are not explained by the past order flow.}
Propagator models were originally proposed in order to solve the so-called \emph{diffusivity puzzle}, namely the fact that price efficiency, and consequently price diffusion, can be achieved even if the order flow imbalances $q_t$ display long-ranged correlation~\cite{Propagator}. Moreover, variations of these models have proven to be effective in order to paint an accurate picture of the market at high frequency \cite{Toth2016,Benzaquen_2017}, in the sense that a large fraction of the price fluctuations can be explained by the past order flow~\cite{Bouchaud_news}.

On the other hand, the perspective taken in order to construct such models is quite distinct from the one preferred in the literature of theoretical economics. 
The propagator setup is not properly microfounded. In fact, it builds on statistical stylized facts, rather than on an economic rationale.
The goal of this paper is to bridge this gap in an economically standard setting by showing how propagator-like models can be rationalized as the  equilibrium resulting from a set of rational agents seeking to achieve  optimality. Along this line, our work is closely related to the classic Kyle setting~\cite{kyle}, in which the price discovery mechanism emerges as a linear equilibrium between three representative agents with asymmetric information. We think that our minimal model, as the Kyle model, is simple enough to be extended in several directions.

We  establish a setting for an Information-Based Model that gives rise to a stationary market, where the equilbrium pricing rule is given by Eq.~\eqref{eq:prop}. A similar setting has been considered in Ref.~\cite{Taub2} in the special case of a stationary Markovian system. Here, instead, we keep the model general, so to account for memory effects (order flows are strongly correlated in real markets), thus extending some of the results of the aformentioned investigation.
Our work goes beyond the purely theoretical aspect,  since the  framework we build allows to explicitly construct kernels $G_t$ that ensure price efficiency under different circumstances.

The organization of the paper is as follows. Section~\ref{sec:notation} introduces the notations we  use throughout the paper. In Section~\ref{sec:model} we present the model. Section~\ref{sec:equilibrium} is devoted to the study of the  equilibrium of the model. Section~\ref{sec:existing_models} discusses the relation of our model with its building blocks, namely the original propagator and the Kyle model. In Section~\ref{sec:Markov} we further investigate the model we propose in the paradigmatic Markovian case, whose tractable solution allows to gain intuition on the system. In Section~\ref{sec:conclusions} we conclude and propose several extensions of our framework.

\section{Notations} \label{sec:notation} Throughout the paper, we will alternate between scalar notations, in which the time dependence of the variables is explicit (e.g. $X_t$), vector notations, and  matrix notations. We will use bold symbols for vectors and Sans Serif symbols for matrices. 

For convenience we introduce two types of vectors: ${\bm X_\past}:= \{X_{t'}\}_{t'=-\infty}^t$ and ${\bm X_{\fut}}:= \{X_{t'}\}_{t'=t}^\infty$.  Further, for a given vector $\bm X_t$ we define the associated Toeplitz matrix as $ \mathsf{X}_{\past,\past} := \{X_{t'-t''}\}_{t',t''=-\infty}^{t}$. In some cases, we will omit the time index for brevity. 
In situations where such omission would be ambiguous, we will restore time indices explicitly, e.g. to deal with matrices such as $\mathsf{X}_{\fut,\fut} = \{X_{t'-t''}\}_{t',t''=t}^{\infty}$ or $\mathsf{X}_{\fut,\past} = \{X_{t'-t''}\}$ with $t' \geq t$ and $t'' \leq t$. The transpose operation will be denoted by the superscript $\top$.

The identity matrix is denoted $\id$, the vector with all components equal to one is written $\vecone$, the matrix with all entries equal to 1 is denoted by $\1$. The Kronecker delta is represented by a vector $\mathbf{e}_t$ with components $e_t(t') = \delta_{t,t'}$. The lag operator, i.e., the operator that acts on an element of a time series to produce the previous element, is denoted $\matL$. In this way we write 
$ \bm{X}_{t-1} = \matL \ \bm{X}_t $. Dimensionless quantities are signified with tildes.

\section{A Simple Agent-Based Market Model}
\label{sec:model}

\subsection{Setup of the Model}

Consider a market in which agents exchange a risky asset (stock) against a safe asset (cash).  The (discounted) transaction price of the risky asset at time $t$ is denoted by $p_t$. Each unit of the risky asset entitles its owner to a stochastic payoff $\mu_t$ in cash (dividend) at each unit of time $t$.
The dividend process $\mu_t$ is modeled as an exogenous, stationary, zero-mean  Gaussian process with autocovariance function (ACF):
\begin{eqnarray}
\label{eq:mu_autocov}
 \Xi^\mu_\tau := \mathbb{E}[\mu_t \mu_{t+\tau}]  \, .
\end{eqnarray}

The portfolio of each agent comprises a combination of risky and safe assets. 
The position of agent $i$ in the risky asset at time $t$ is given by $Q^i_t$, whereas his trades are denoted by $q^i_t := Q^i_{t} - Q^i_{t-1}$.
With these conventions, the equations for the evolution of cash $C^i_t$, stock-position  $Q^i_t$, and wealth $W^i_t$ for each
agent can be written down respectively as: 
\begin{eqnarray}
    \label{eq:risk_neutral_utility}
    \Delta C^i_t &:=&  \mu_{t} Q^i_t - p_t q^i_t \\
    \Delta Q^i_t &:=& q^i_t\, 
    \\
    \Delta W^i_t &:=& \Delta C^i_t + Q^i_t p_t - Q^i_{t-1} p_{t-1}.
\end{eqnarray}

We consider an agent-based market model with asymmetric information akin to the well known Kyle model \cite{kyle}, in which the agents take actions at discrete time steps $t$. A strategic agent possessing privileged information about the realizations of the stochastic dividend process (\emph{informed trader}, or IT) trades with a non-strategic and non-informed trader (\emph{noise trader}, or NT) that accesses the market for exogenous reasons. Both the IT and NT are modeled as liquidity takers. A liquidity provider (\emph{market maker}, or  MM) provides liquidity for both the NT and the IT and sets the transaction price $p_t$.

At the beginning of each time interval $[t,t+1]$ both the IT and the NT build a demand for the risky stock $q^i_t$ (with $i \in \{\IT,\NT\}$). \emph{After} the excess demand $q_t: = q^{\IT}_t + q^{\NT}_t $ is formed, the MM clears the excess demand of the liquidity takers, executing a trade $q^{\mathrm{MM}}_t := -q_t$ and setting the transaction price  $p_t$. The  price~$p_t$ arises endogenously as the result of the action of the agents, described in what follows.

Before discussing the information sets and the strategies of the different agents, let us highlight that both the IT and the MM have exact knowledge of the statistical properties of the exogenous processes $\mu_t$ and $q^\NT_t$, as well as each other's strategy. Past prices and excess demands are also public information.

\paragraph{Noise trader} The NT acts in a purely stochastic fashion. His demand process $q^\NT_t$ is a zero-mean, stationary Gaussian process with ACF given by:
\begin{equation}
\label{eq:CN_autocov}
\autocvol^{\NT}_\tau:= \avg[q^\NT_t q^\NT_{t+\tau}]  \, .
\end{equation}

\paragraph{Informed trader} The IT is a strategic, risk-neutral (expected)  utility maximizer.   His access to privileged information about the dividend process is modeled by assuming that he observes past  realizations of the process $\mu_t$ and uses such information to maximize his future expected wealth.
Moreover, since realized past excess demand is public information, the IT  can trivially infer the NT's  past trades. 
The information accessible to the IT at time $t$ is thus given by: 
\begin{equation}
\label{eq:IT_info}
\filt^{\IT}_\past = \left\{  \mathbf{q}_{\past-1},\mathbf{q}^{\NT}_{\past-1}, \boldsymbol{\mu}_{\past-1} \right\}.     
\end{equation}
 So the IT builds his demand without exploiting equal-time information on either $p_t$ nor on the decision of his peers. 
In order to maximize his wealth, the IT exploits privileged information on realized  dividends.

Since the IT is risk-neutral and  assuming that the price is a  linear function of realized excess demands (we shall discuss why this is  the case in a moment), his demand \(q^{\IT}_t\) at time \(t\)  is a \emph{linear} function of his current information set  \(\filt^{\IT}_t\): 
\begin{eqnarray}
    \label{eq:demand_kernel}
    \mathbf{q}^\IT_\past 
    &=& 
     \matdmd \mathbf{q}_{\past-1}+ \matdmd^\NT \mathbf{q}^{\NT}_{\past-1}+ \mathsf{R}^{\mu} \boldsymbol{\mu}_{\past-1},
\end{eqnarray} 
where we have introduced the
 \emph{demand}  kernels ($\matdmd,\matdmd^\NT,\mathsf{R}^\mu$). 
Let us give here a first description of these demand kernels. Since we discuss a market with multiple trading periods, the IT strategically takes into account past trades and past dividends in order to determine his demand. 
The demand kernel $\matdmd$ accounts for the dependence on past order flow which arises from price impact of past traded volumes. The kernel $\matdmd^{{\NT}}$ accounts for the dependence that comes from the price impact induced by expected future trades of the NT, while the kernel $\matdmd^\mu$ accounts for the dependence arising from expected future dividends. The demand kernels are the result of a Model Predictive Control (MPC)~\cite{MPC} strategy. Indeed, as soon as a new piece of information is available to the IT (i.e. at each time-step $t$), he will construct an updated long-term strategy, and he will trade accordingly.
More details about the IT's MPC strategy are provided in  Sec.~\ref{sec:opt_trading}, with explicit expressions of the demand kernels.

\paragraph{Market maker} 
The MM is risk-neutral and competitive. He  sets a pricing rule that allows him to statistically break even on every trade, without controlling the inventory that he might accumulate while matching the demand.
The past realization of the dividend process $\mu_t$ is unknown to the MM, and so is the proportion of the demand due respectively to the IT and the NT. Thus, the information set available to the MM at time $t$ is solely given by realized aggregate excess demand:  
\begin{equation}
    \filt^{\MM}_t :=     \{\mathbf{q}_\past\}.
\end{equation} 
An important point is that the resulting excess demand $q_t$  conveys  information to the MM about the asset's fundamental value, via the information set used by the IT (Eq.~\eqref{eq:IT_info}) to construct his trading schedule (Eq.~\eqref{eq:demand_kernel}). Note also that the information set of the MM is \emph{not} contained in the information set of the IT, due to the fact that the excess demand $q_t$ is only available to the IT at time $t+1$.

 Since the MM knows that the IT' s trading schedule is given by Eq.~\eqref{eq:demand_kernel}, from the total order flow he can infer information about past dividends, albeit this information is distorted by the presence of noise induced by the NT.  From Eq.~\eqref{eq:demand_kernel}, the dynamics of the excess demand at time is linear in NT's trades up to time $t$ and past dividends and it is given  by:
\begin{equation}
\label{eq:excess_demand_dynamics}
    \mathbf{q}_\past = \left(\id-\matdmd \mathsf{L}\right)^{-1}\left[\left(\id+\matdmd^{\NT} \matL \right)\mathbf{q}^{\NT}_{\past} + \matdmd^{\mu} \boldsymbol{\mu}_{\past-1} \right].
\end{equation} 

Due to the Gaussian nature of both $\mu_t$ and $q^\NT_t$ and the risk-neutral nature of  market participants, the  choice of considering a linear (instead of a general) equilibrium implied by Eq.\eqref{eq:prop} appears natural. On the other hand, we do not have a proof of uniqueness of the linear stationary equilibrium. Given that such uniqueness holds in a framework similar to ours (given by Ref.~\cite{proof_uniqueness}), it is reasonable to expect that uniqueness should hold even in our case. Thus, the market can be modeled by the MM as a Linear Gaussian State-Space Model (LG-SSM)~\cite{Murphy}. Actually, while the state of the market,  i.e. realized dividends $\boldsymbol{\mu}_\past$ and NT's trades $\mathbf{q}_\past$,  are not observable by the MM, he can infer these quantities, and in particular realized dividends, filtering them out from his information set.
 This procedure in the LG-SSM literature is referred to as Kalman filtering technique. More details about these important aspects of the model will be given in the following section.  

\subsection{Competitive pricing rule}
\label{sec:pricing_rule}
As anticipated above, we assume the MM to be competitive and risk neutral. Thus, by a Bertrand auction type of argument~\cite{kyle}, we postulate a break even condition for the MM for each 
 $T$-period holding strategy built as follows: buy \(q_t\) units of stock by matching the demand at time $t$ at a price \(p_t\) and sell them back at time \(t+T\) at a price  $\avg[p_{t+T}|\filt^\MM_t]$, earning the dividends in the meanwhile. Note that even though the MM cannot choose to execute with certainty at $t+T$, we can see $T$ as the time lag at which the MM decides to mark-to-market his position, even if he might not be actually able to liquidate it. 
Imposing competitiveness of the MM, this trajectory should have zero payoff on average, leading us to postulate a pricing-rule of the form:
\begin{equation}
    \label{eq:pricing_rule}
    p_t =  \sum_{t'=t}^{t + T - 1} \avg[ \mu_{t'} | \filt^\MM_t ] + \avg[p_{t+T}| \filt^\MM_t] . 
\end{equation}
Thus, the price at time $t$ is given by the long-term sum of future dividends plus a boundary term which in general is non-zero. 

    \subsubsection*{Stationary dividends with zero mean}
    
    If the boundary term in  Eq.~\eqref{eq:pricing_rule} evaluated at $T=\infty$ is equal to zero, i.e., the transversality condition holds, one obtains the standard EMH fundamental rational expectation  pricing-rule:
    \begin{equation}
      \label{eq:myopic}
         p_t = \avg \left[ p^\text{F}_t | \filt^\MM_t \right],  \,  \ \  \text{where} \ \  p^\text{F}_t = \boldsymbol{1}_{\fut}^\top \  \boldsymbol{\mu}_{\fut}.
    \end{equation}
    In case of mean-reverting  dividends process with zero mean, the transversality condition is justified. We will investigate the model with this assumption, for simplicity reasons. Under this prescription the job of the MM is to provide the optimal forecast of discounted future cash flows from infinity to the present time $t$, given his current information set. Notice that restoring a fundamental price with non-zero mean would simply amounts to a rigid (although, infinite) shift of the price process, since the mean of the fundamental price is public information and so it is immediately incorporated into the price.
    
    It will be interesting to compare the result of the MM's estimate, given by Eq.~\eqref{eq:myopic}, with the  one constructed by the IT, which is not distorted by the noise induced by the NT:
    \begin{equation}
    \label{eq:p_best}
    p^{\IT}_t = 
    \avg\left[\left.p^\text{F}_t \right|\filt^{\IT}_t\right].
    \end{equation}
Let us note here that the dividends have to be predictable for the market to be non trivial. In fact, if the dividend process is not correlated, i.e.,  $\Xi^\mu_\tau = \Xi^\mu_0 \delta_\tau$, then $p^\IT_t = 0$, i.e., the IT does not have any informational advantage over the MM. Thus, in this case, the MM would simply set the price equal to zero.

With the pricing rule given by Eq.~\eqref{eq:myopic} the  MM statistically breaks even for each buy or sell trade, if he waits enough time for the income due to the dividends to restore his cash account to zero. This local constraint is thus given by:
\begin{equation}
\label{eq:avg_break_even}
     \avg[\Delta C_t^\MM] = 0.
\end{equation}
As a consequence  $\avg[\Delta C^\IT_t]+\avg[\Delta C^\NT_t ]= 0$, i.e. the gain of the IT is balanced by the losses of the NT. This is what typically happens in models where NT are uninformed and non-rational~\cite{O'Harabook}.  

In the following we give the explicit expression of the pricing rule \eqref{eq:myopic} in terms of the IT's trading schedule, i.e. in terms of the IT's demand kernels introduced in Eq.~\eqref{eq:demand_kernel}.

\subsubsection*{Dividends  regression  from observed excess demand}
\label{subsubsec:filtering}
 
The pricing rule given by Eq.~\eqref{eq:myopic} prescribes that the MM should estimate the sum of future dividends by observing realized excess demand. This problem can be solved in two steps. First the MM estimates realized dividends applying a filter on realized excess demand. The optimal estimator of realized dividends is well known in the LG-SSM literature as Kalman filter and it is linear in the measurements, i.e., the realized excess demand in our model. Then, the MM computes the expected sum of future dividends summing over the forecasts of future dividends.  In the following we detail these two steps.

The MM's estimate of realized dividends $\hat{\boldsymbol{\mu}}_\past:= \avg[\boldsymbol{\mu}_\past|\filt^\MM_t]$ is given by $\hat{\boldsymbol{\mu}}_\past  = \mathsf{K} \mathbf{q}_\past$, 
where we have implicitly defined the (steady-state) Kalman gain $\mathsf{K}$. This matrix can be constructed in a standard way \cite{Murphy, Kalman} given  the dynamics of the MM's measurements, i.e.,  Eq.~\eqref{eq:excess_demand_dynamics}. 
The Kalman gain $\mathsf{K}$ is proportional to the signal noise, i.e., $\Xi^\mu$, and inversely proportional to the measurement noise, which is the ACF of the excess demand  $\Omega_{\tau}: = \avg[q_t q_{t+\tau}]$ and it is explicitly given by:
\begin{equation}
\label{eq:Kalman}
    \mathsf{K} = \mathsf{\Xi^\mu} (\mathsf{J}^\mu)^\top \mathsf{\Omega}^{-1},
\end{equation}
where\footnote{ Using the Woodbury identity on Eq.~\eqref{eq:Kalman}, one obtains the alternative  expression of the  gain matrix $\mathsf{K}$:
\begin{equation}\nonumber
\label{eq:div_filter}
    \mathsf{K} = \left[
\left(\mathsf{\Xi^\mu}\right)^{-1} + \left(\mathsf{J}^\mu \right)^\top  \left(\mathsf{D}^\NT\right)^{-1}\mathsf{J}^\mu
\right]^{-1} \left(\mathsf{J}^\mu \right)^\top \left(\mathsf{D}^\NT\right)^{-1}.  
\end{equation}
This alternative expression gives a complementary interpretation of the gain matrix $\mathsf{K}$: in fact the  matrix inside the square bracket is the dividends posterior information matrix. This matrix is given by the dividends prior information matrix $\left(\mathsf{\Xi^\mu}\right)^{-1}$ summed to the information added by the measurement, i.e., $\left(\mathsf{J}^\mu\right)^\top  \left(\mathsf{D}^\NT\right)^{-1}\mathsf{J}^\mu$. 
}
\begin{equation}
\begin{split}
    \mathsf{J}^\mu &= \left( \id -\matdmd \matL \right)^{-1} \matdmd^\mu \matL
    \\
    \mathsf{\Omega} &= \mathsf{J}^\mu   \mathsf{\Xi^\mu}\left(\mathsf{J}^\mu\right)^\top + \mathsf{D}^\NT.
\end{split}
\end{equation}
$\mathsf{J}^\mu$ is the matrix that multiplies the dividends in the r.h.s. of Eq.~\eqref{eq:excess_demand_dynamics} and $\mathsf{D}^\NT$ is the NT's dressed ACF, given by:
\begin{equation}
    \mathsf{D}^\NT = \left(\id-\matdmd \matL \right)^{-1} \left(\id+\matdmd^\NT \matL \right) \mathsf{\autocvol}^\NT \left(\id+\matdmd^\NT \matL \right)^\top\left[\left(\id-\matdmd \matL \right)^{-1}\right]^\top.
\end{equation}
 The noise ACF is dressed since the noise (i.e., the NT's trade process) not only affects the excess demand dynamics by construction ($\mathbf{q}_t = \mathbf{q}^\IT_t+\mathbf{q}^\NT_t $), but also because the  IT's optimal trading strategy depends upon past and future realizations of the noise (see  Eq.~\eqref{eq:demand_kernel}).

From estimated realized dividends $\hat{\boldsymbol{\mu}}_t$, the MM has to estimate the fundamental price $p^\text{F}_t$, defined in Eq.~\eqref{eq:myopic}. To do so, he builds the forecast of future dividends as $
    \avg[\boldsymbol{\mu}_{\fut}| \hat{\boldsymbol{\mu}}_\past] = \mathsf{F}^\mu \hat{\boldsymbol{\mu}}_\past$,
where we introduced
the dividends forecast matrix \(\mathsf{F}^\mu\). Since the dividends process is Gaussian with zero-mean, \(\mathsf{F}^\mu\) depends only on the ACF of the dividends $\mathsf{\Xi^\mu}$.
Finally, by summing over the estimated future dividends we obtain the following equation for the price at time $t$:
\begin{equation}
\label{eq:myopic_expanded}
    p_t = \boldsymbol{1}_{\fut}^\top  \ \mathsf{F}^\mu \mathsf{K} \ \mathbf{q}_\past.
\end{equation}
Notice, that Eq.~\eqref{eq:myopic_expanded} explicitly gives the rule for the propagator $\mathsf{G}$. In fact, in compact notation, the propagator model is given by $p_t = \boldsymbol{1}_{\fut}^\top  \ \mathsf{G} \ \mathbf{q}_\past$\footnote{Compare with Eq.~\eqref{eq:prop}, where the propagator model is introduced.}.

In the following section we construct the IT's optimal trading strategy based on the maximization of his expected future wealth, as a function of the MM's pricing rule. This means that, as anticipated, the IT's demand kernels ($\matdmd, \matdmd^\NT,\matdmd^\mu$) are functions of the propagator $\mathsf{G}$ introduced in Eq.~\eqref{eq:prop}, and so is the Kalman gain matrix $\mathsf{K}$ introduced in Eq.~\eqref{eq:div_filter}. Because of this, Eq.~\eqref{eq:myopic_expanded} will turn out to be a self-consistent equation for the propagator~$\mathsf{G}$.

\subsection{Optimal insider trading}

\label{sec:opt_trading}

The utility function $U^\IT_t$, whose expectation is  maximized by the IT at each time step $t$, is defined by the value of his wealth account at a terminal time $t+T$ (where $T$ is not related to that introduced in Sec.~\ref{sec:pricing_rule}), given by $W^\IT_{t+T}$, in which his position $Q^\IT_t$ in the risky asset is flattened. Thus, $\util^\IT_t = W^\IT_{t+T}$ subject to the constraint $Q^\IT_{t'} = 0$ for $t' \geq t + T$.

At each time step $t$, the IT optimizes his expected utility function  over the whole future trajectory $\mathbf{q}^{\IT}_{\fut}$ given the information set at the current time $\filt^{\IT}_t$ given by  Eq.~\eqref{eq:IT_info}, and trades the first step of the optimal strategy. The IT's trade at time $t$ is thus calculated as follows:
\begin{equation}
\label{eq:IT_strategy}
    q^{\IT}_t = \mathbf{e}_t^\top \argmax_{\mathbf{q}^{\IT}_{\fut}} \mathbb{E}\left[\left.\util^{\IT}_{t} \right| \filt^{\IT}_t \right],
\end{equation}
where  $\mathbf{e}^\top_t$ explicits the fact that only the first step of the future trajectory is executed.
Notice that the presence of a finite liquidation time does not break the assumption of the time-translational invariance of the model, because the terminal condition is also receding as time moves on.
Indeed, the IT will in general hold a non-zero position $Q^\IT_t$ up to $t\to\infty$ despite the presence of the liquidation constraint. The constraint should then be seen as a device used by the IT in order to properly mark-to-market the value of his \emph{current} stock positions at time $t$ by taking into account the forecast of their \emph{future} liquidation value $p_{t+T}$, rather than as a measure taken to prevent him from trading at large times.

In the following we analyze  the case in which  \(T=\infty\) with  mean-reverting dividends.

\subsubsection*{Stationary demand kernels of the insider with infinite horizon}

If $T = \infty$ in Eq.~\eqref{eq:IT_strategy}, the IT can neglect the round-trip constraint, since liquidation costs are pushed to the far away future and, due to the assumptions of zero-mean and mean-reverting dividends, the expected price at infinity is  zero. Because of this, the actual trading profile of the IT that we will consider in the following is given  by Eq.~\eqref{eq:IT_strategy} with $U^{\IT}_{t} = C^{\IT}_\infty$. 
In doing so, the maximization program is given by
\begin{equation}
    \label{eq:gain_infinity}
    q^{\IT}_t = \mathbf{e}_t^\top \argmax_{\mathbf{q}^{\IT}_{\fut}} \mathbb{E}\left[\left. C^\IT_\infty \right| \filt^{\IT}_t \right], \quad \text{where} \quad C^\IT_\infty = C^\IT_{t-1} -  \left(\mathbf{q}^\IT_{\fut}\right)^\top\left( \mathbf{p}_{\fut}-\mathbf{p}^\text{F}_{\fut}\right).  
\end{equation} 
 In order to keep the discussion simple we  consider the dividend process with integrable ACF, such that the IT's estimate of the fundamental price $p^\text{F}_t$ is finite. One can in fact relax this hypothesis, with a suitable renormalization of the price and dividends process.

Notice that the introduction of a non-zero mean for the fundamental price does not affects nor the IT's strategy nor the price impact function. In fact, since the expectation of the fundamental price is assumed to be public information, the MM could immediately incorporate it in the price, as discussed previously. Then, since the IT's gain in Eq.~\eqref{eq:gain_infinity} is proportional to the difference between the price and the IT's estimate of the fundamental price, it follows that the IT's trading strategy does not depend on the mean of the fundamental price. To conclude, since the propagator given by Eq.~\eqref{eq:myopic_expanded}  depends only on IT's demand kernels ($\matdmd, \matdmd^\NT,\matdmd^\mu$) and ACFs of dividends and NT's trades $(\Xi^\mu,\Omega^\NT)$ via the Kalman filter (Eq.~\eqref{eq:Kalman}) and the dividend forecast matrix $\mathsf{F}^\mu$, it follows that the mean of the fundamental price is immaterial in shaping the the price impact function.

The expression for the demand kernels ($\matdmd, \matdmd^\NT,\matdmd^\mu$) at equilibrium can be determined as solution of the quadratic optimization program defined by  Eq.~\eqref{eq:gain_infinity}.  
The expected gain at infinity $C^\IT_\infty$ depends on estimated future dividends (via $\mathbf{p}^\text{F}_{\fut}$) and on  estimated future NT's trades (via $\mathbf{p}_{\fut}$). Thus, in order to write it down explicitly,  we need the dividends forecast matrix $\mathsf{F}^\mu$ introduced in the previous section, and the forecast matrix of NT's trades, $\mathsf{F}^{{\NT}}$, defined similarly by $
    \avg[\mathbf{q}^{\NT}_{\fut} |\mathbf{q}^{\NT}_\past]
    = 
     \matfcast^{{\NT}}  \mathbf{q}^\NT_\past
$. 

Since $\avg[C^\IT_\infty|\filt^{\IT}_\past]$ depends on past realizations and  forecasts, we insert  time subscripts over matrix symbols in order to avoid ambiguities. We  obtain: 
\begin{equation}
\label{eq:gain_infty_exp}
\begin{split}
    \avg[C^{\IT}_\infty|\filt^{\IT}_t] =& -\frac{1}{2} \left(\mathbf{q}^\IT_{\fut}\right)^\top \matsymprop_{\fut,\fut} \mathbf{q}^\IT_{\fut}
    \\ &-  \left(\mathbf{q}^\IT_{\fut}\right)^\top \left[ \mathsf{G}_{\fut,\past-1}\mathbf{q}_{\past-1}+ \mathsf{G}_{\fut,\fut}\mathsf{F}^{{\NT}}_{\fut,\past-1}\mathbf{q}^{\NT}_{\past-1} - \1_{\fut,\fut} \matfcast^\mu_{\fut,\past-1} \boldsymbol{\mu}_{\past-1}\right],
\end{split}
\end{equation}
where we dropped $C^{\IT}_{t-1}$, since it does not depend on IT's future trades  $\mathbf{q}^{\IT}_{\fut}$, and we introduced the symmetric propagator $\matsymprop = (\matprop + \matprop^\top)$ in order to write in a compact form the quadratic term in  $\mathbf{q}^\IT_\fut$.  The quadratic term in $\mathbf{q}^\IT_\fut$  of Eq.~\eqref{eq:gain_infty_exp} is the cost term that the IT will face due to his own future market impact, while the the linear term in $\mathbf{q}^\IT_{\fut}$ is his signal term. The first term of the signal comes from price impact due to known order flow realizations, the second one comes  from the expected price impact of future NT's trades, while the third one comes from his private information about $\mathbf{p}^\text{F}_\fut$. 

The expression for the IT's demand kernels, defined in Eq.~\eqref{eq:demand_kernel}, can be obtained in terms of the propagator $\mathsf{G}$ and the forecast matrices $\mathsf{F}^\NT$ and $\mathsf{F}^\mu$ inserting Eq.~\eqref{eq:gain_infty_exp} in Eq.~\eqref{eq:gain_infinity}:
\begin{subeqnarray}
\label{eq:IT_response}
    \vecdmd_t &=& - \mathbf{e}^\top_{t} \left[\matsymprop_{\fut,\fut}\right]^{-1}  \matprop_{\fut,\past-1},
    \\
\label{eq:R_NT}
    \vecdmd^{{\NT}}_t &=& -\mathbf{e}^\top_t\left[\matsymprop_{\fut,\fut}\right]^{-1} \matprop_{\fut,\fut} \matfcast^{{\NT}}_{\fut,\past-1},
    \\
    \vecdmd^\mu_t &=& \mathbf{e}^\top_t \left[\matsymprop_{\fut,\fut}\right]^{-1} \1_{\fut,\fut} \matfcast^\mu_{\fut,\past-1}.
\end{subeqnarray}

Finally, we have all the ingredient to write down explicitly the functional equation for the equilibrium pricing rule, which will be given in the following section.

\section{The linear equilibrium }
\label{sec:equilibrium}

\subsection{Equilibrium condition and numerical solution}
The linear equilibrium of the model can be found by self-consistently taking into account the competitive pricing rule of the MM and the  strategy of the IT, given respectively in   Eqs.~\eqref{eq:myopic_expanded} and \eqref{eq:IT_response}. 
The self-consistent  equation for the propagator, in scalar notation,  reads:
 \begin{equation}
\label{eq:master_equation}
G_{t-s} = \sum_{t'=t}^\infty \sum_{t''=-\infty}^t F^\mu_{t',t''} {K}_{t'',s}[G]
\end{equation}
where we have made it explicit that the filter $K$ is a function of the propagator $G$ itself: in fact, $K$ is  given in terms of IT's demand kernels ($\matdmd, \matdmd^\NT,\matdmd^\mu$) by Eq.~\eqref{eq:Kalman}, which depend on the propagator $G$ via Eqs.~\eqref{eq:IT_response}.

The linear equilibrium equation \eqref{eq:master_equation} is a non-linear functional equation for the propagator $G_{t}$. 
As such it is not amenable for analytical treatment  in the general case of arbitrary Gaussian, zero-mean and stationary dividends and NT's trades process. 
Nevertheless, we have been able to solve Eq.~\eqref{eq:master_equation} iteratively, as illustrated in Appendix~\ref{sec:numerical_solver}.
In two special  cases  we have been able to validate the result of the iterative numerical solver by means of the analytical solution of  Eq.~\eqref{eq:master_equation}  (see Appendix~\ref{app:solution_master}).

Via an extensive analysis of the model based on the iterative numerical solver of Eq.~\eqref{eq:master_equation} we found that the market at equilibrium exhibits some robust properties, that hold in case of an integrable and stationary ACF  of the NT's trades and dividends, regardless of the exact structure of the ACFs.
These properties are listed below.

\subsection{Generic equilibrium properties}
\label{sec:general_prop}

\paragraph{Return covariance}
     The equilibrium is characterized by a return  ACF  $\Xi_\tau := \avg[\Delta p_t \Delta p_{t+\tau}]$ with the same temporal structure as that related to the IT's price estimate $p^\IT_t$,  given by Eq.~\eqref{eq:p_best}, which will be referred to as $\Xi^\IT_\tau$. In formula: \begin{equation}
    \label{eq:price_eff}
        \Xi_\tau = \Xi_0   \tilde{\Xi}^\IT_\tau, \ \text{with} \ \tilde\Xi^\IT_0 = 1.
    \end{equation}
     The price distortion induced by the noise injected into the system by the NT is thus  completely encoded in a scalar, the return variance  \(\Xi_0\).   
    
\begin{figure}
    \centering
    \includegraphics[scale = 0.5]{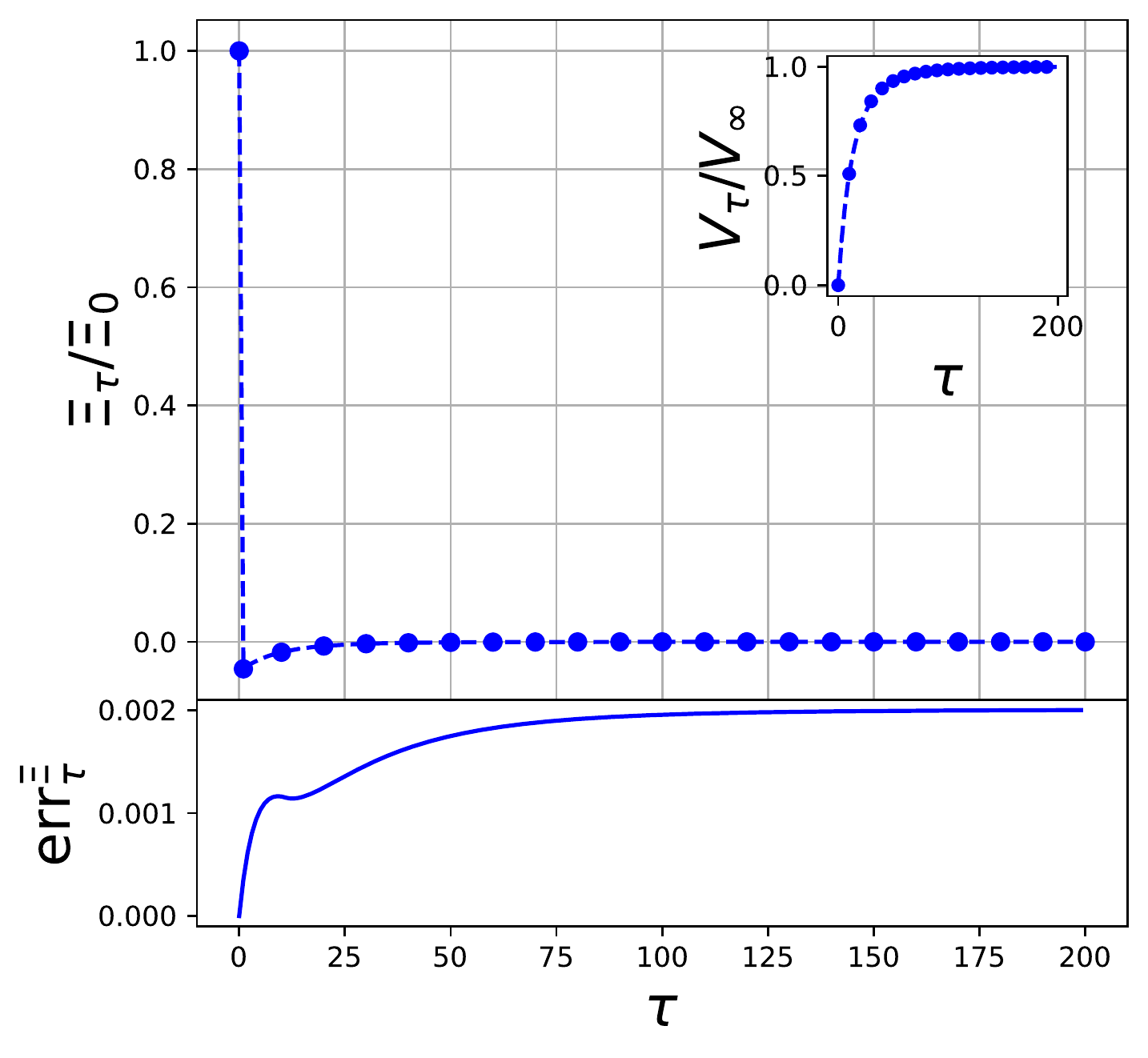}
    \hspace{1cm}
    \includegraphics[scale = 0.5]{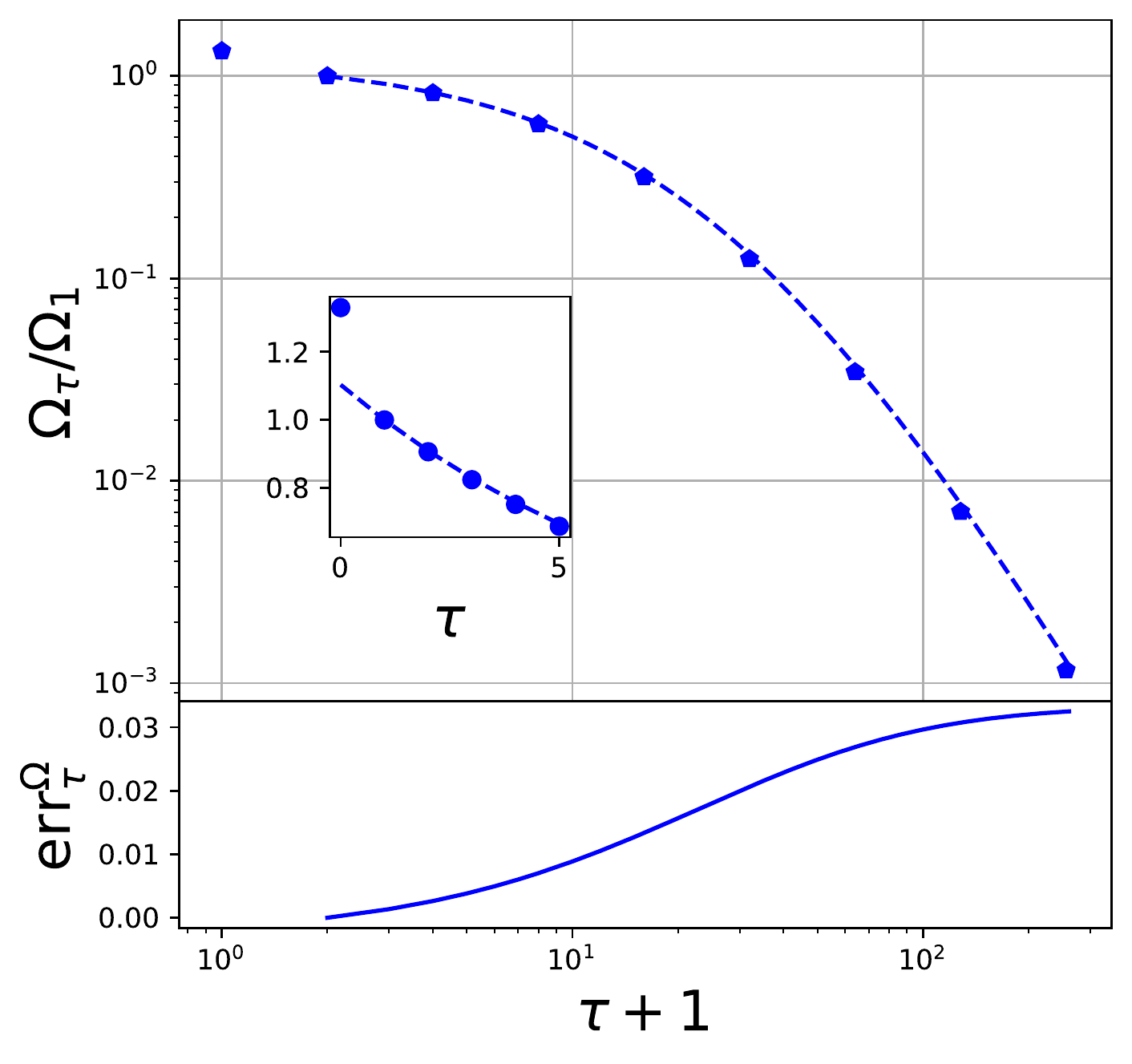}

    \caption{
     Numerical check of equilibrium properties with  ACFs given by $(1+|\tau|/\tau_{k})^{-\gamma_k}$ where $k = \{\mu,\NT\}$. We arbitrarily choose $(\tau_\NT, \tau_{\mu} ,  \gamma_\NT , \gamma_\mu ) = (30,50,3,5)$. The numerical solver has been implemented with $T_{cut} = 5 \cdot 10^2$ and $T_{it} = 200$. (Left) In the upper panel we show the good collapse between  $\Xi_\tau/\Xi_0$ (bullet points) and  $ \Xi_\tau^\IT/\Xi_0^\IT$ (dashed line). The collapse between these two ACFs  is quantified  in the bottom panel, where the relative cumulative absolute error between the two curves is displayed.  The inset in the top panel shows the collapse on the variogram. (Right) In the main top panel we show the good collapse for positive lags between  $ \autocvol_\tau/\autocvol_1$ (bullet points) and $ \autocvol_\tau^\NT/\autocvol_1^\NT$ (dashed line), whereas in the inset we show that the collapse doesn't involve the lag 0 term.
    In the bottom panel the collapse between these two ACFs is quantified, calculating the relative cumulative absolute error starting from lag 1. }
    \label{fig:example_pl}
\end{figure}

\begin{figure}
    \centering
    \includegraphics[scale = 0.5]{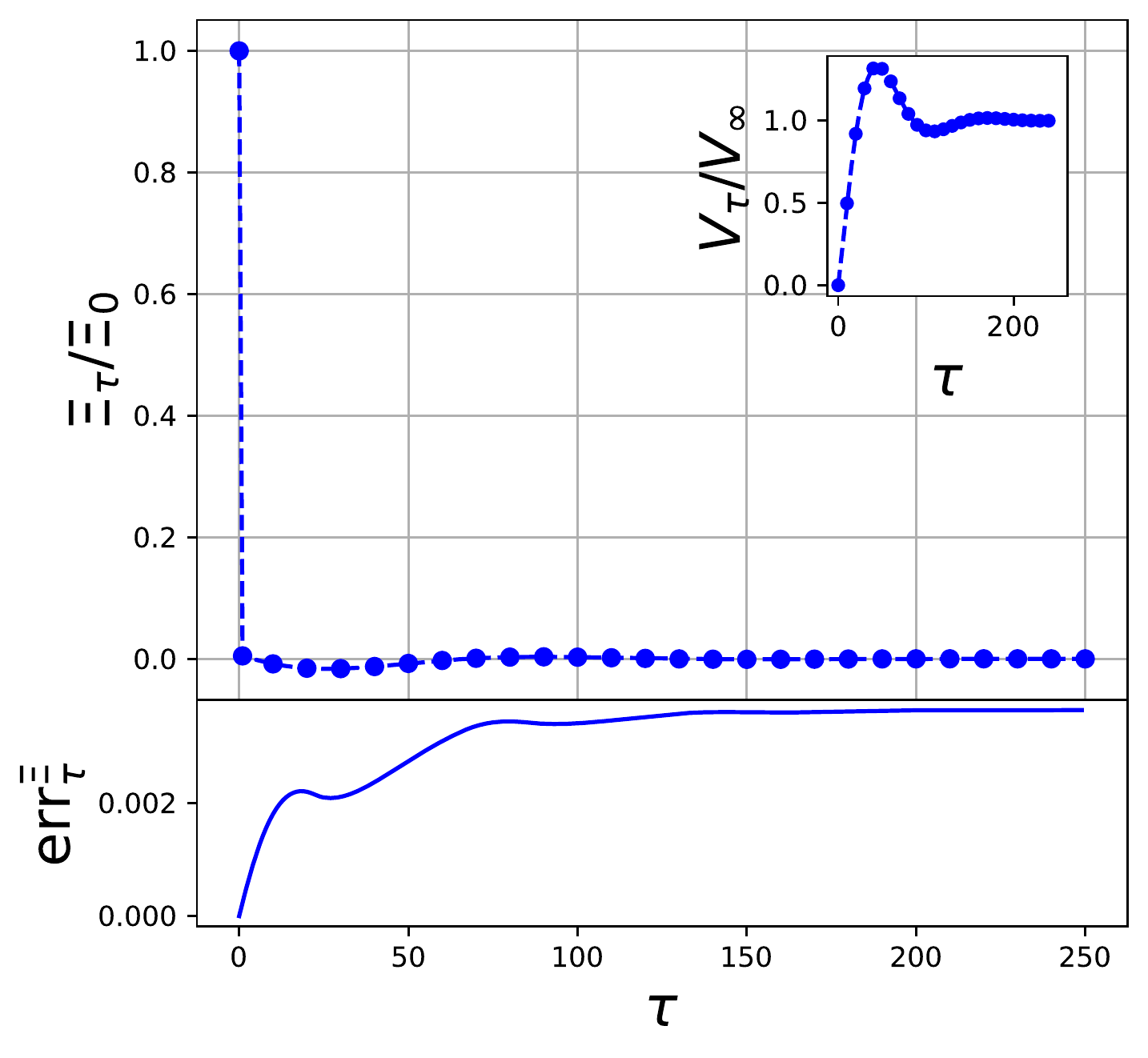}
    \hspace{1cm}
    \includegraphics[scale = 0.5]{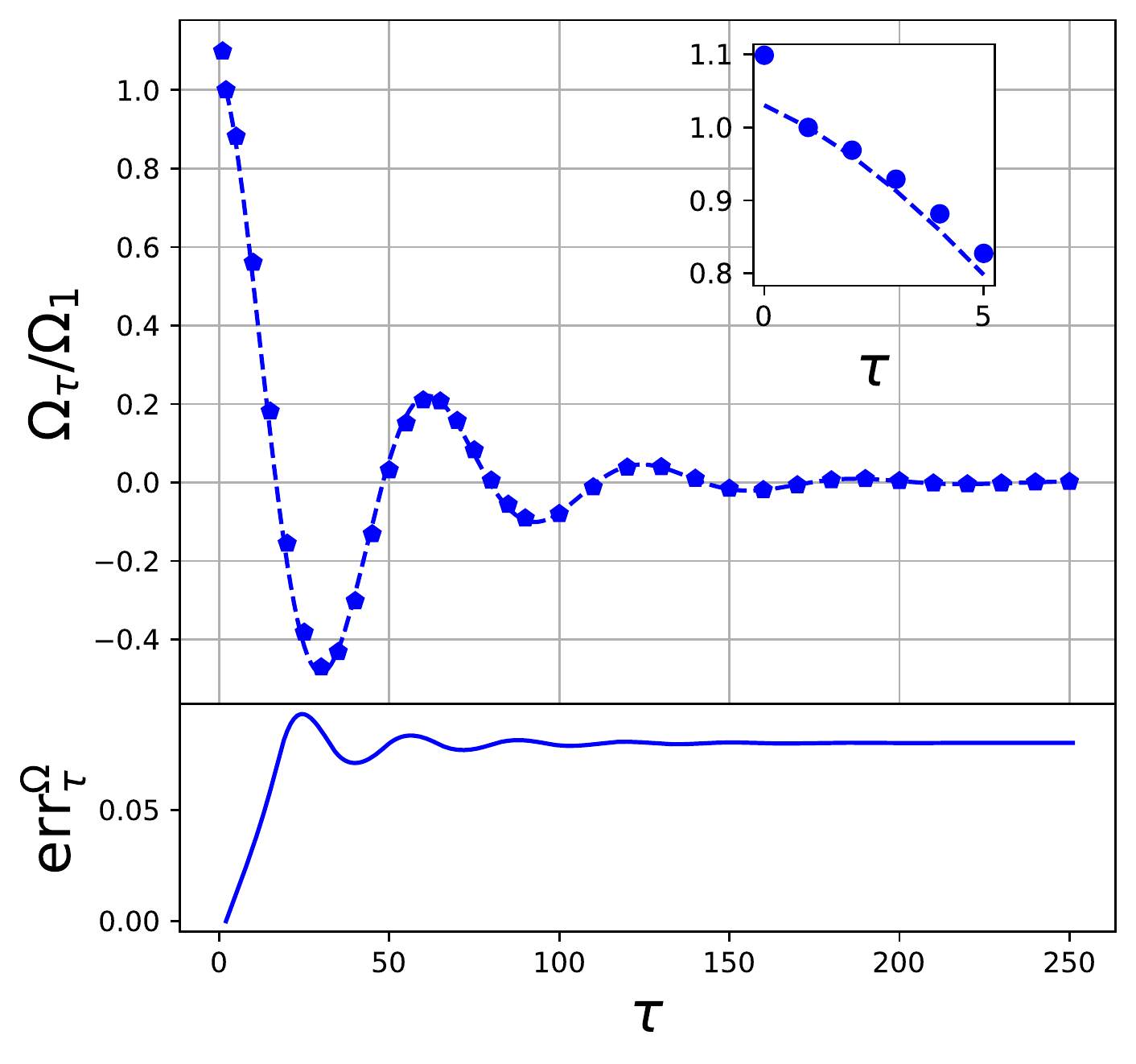}

    \caption{
     Numerical check of equilibrium properties with  ACFs given by $\exp^{-\tau/\tau_{1,k}} \sin(x/\tau_{2,k} +\pi/2)$ where $k = \{\mu,\NT\}$. We arbitrarily choose
     $(\tau_{1,\NT},\tau_{1,\mu},\tau_{2,\NT}, \tau_{2,\mu}) = (40,40,20,10)$. The numerical solver has been implemented with $T_{cut} =  10^3$ and $T_{it} = 500$. (Left) In the upper panel we show the good collapse between  $\Xi_\tau/\Xi_0$ (bullet points) and  $ \Xi_\tau^\IT/\Xi_0^\IT$ (dashed line). The collapse between these two ACFs  is quantified  in the bottom panel, where the relative cumulative absolute error between the two curves is displayed.  The inset in the top panel shows the collapse on the variogram. (Right) In the main top panel we show the good collapse for positive lags between  $ \autocvol_\tau/\autocvol_1$ (bullet points) and $ \autocvol_\tau^\NT/\autocvol_1^\NT$ (dashed line), whereas in the inset we show that the collapse doesn't involve the lag 0 term.
    In the bottom panel the collapse betweeen these two ACFs is quantified, calculating the relative cumulative absolute error starting from lag 1. }
    \label{fig:example_osc}
\end{figure}

\begin{figure}
    \centering
    \includegraphics[scale = 0.5]{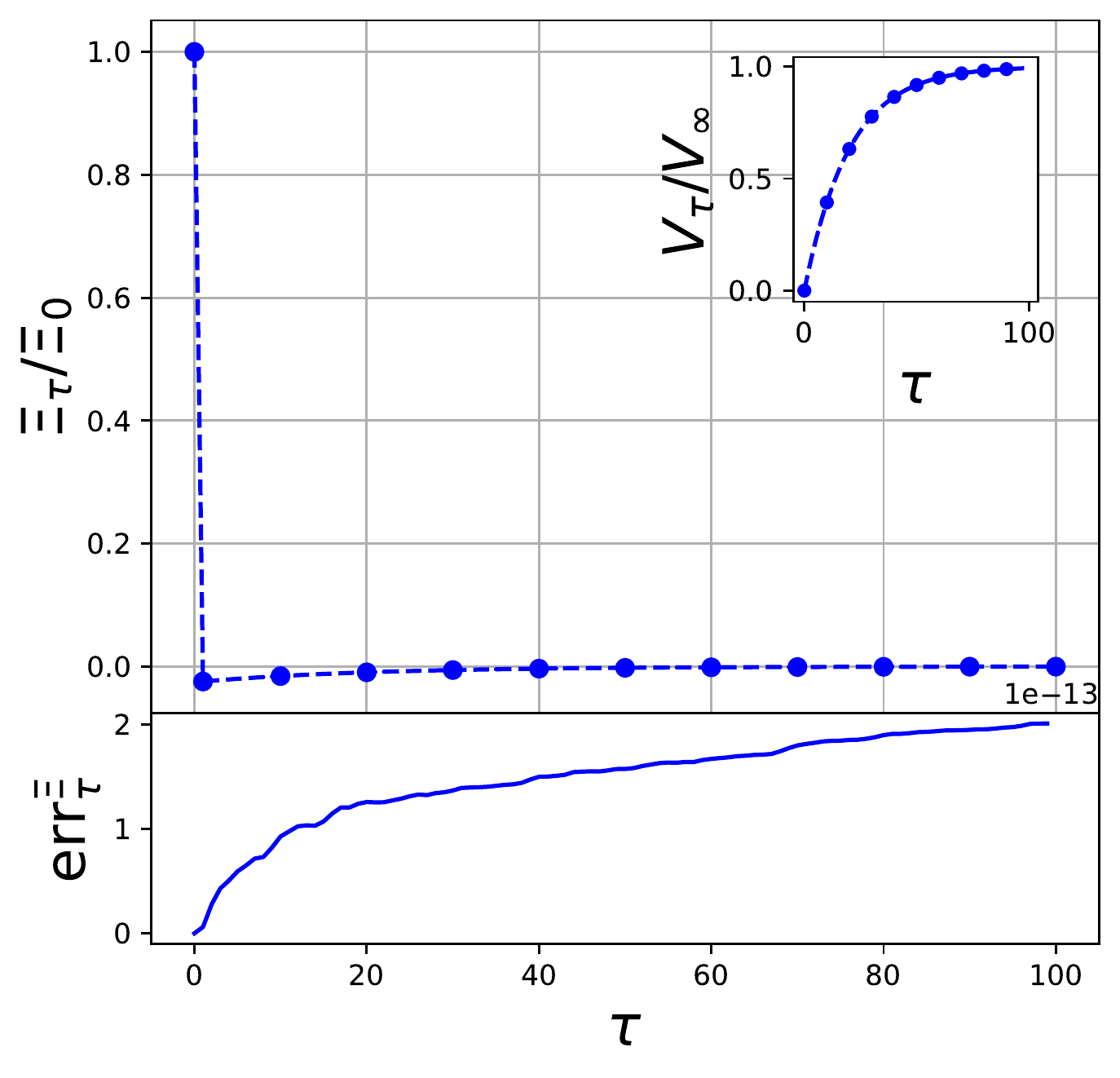}
    \hspace{1cm}
    \includegraphics[scale = 0.5]{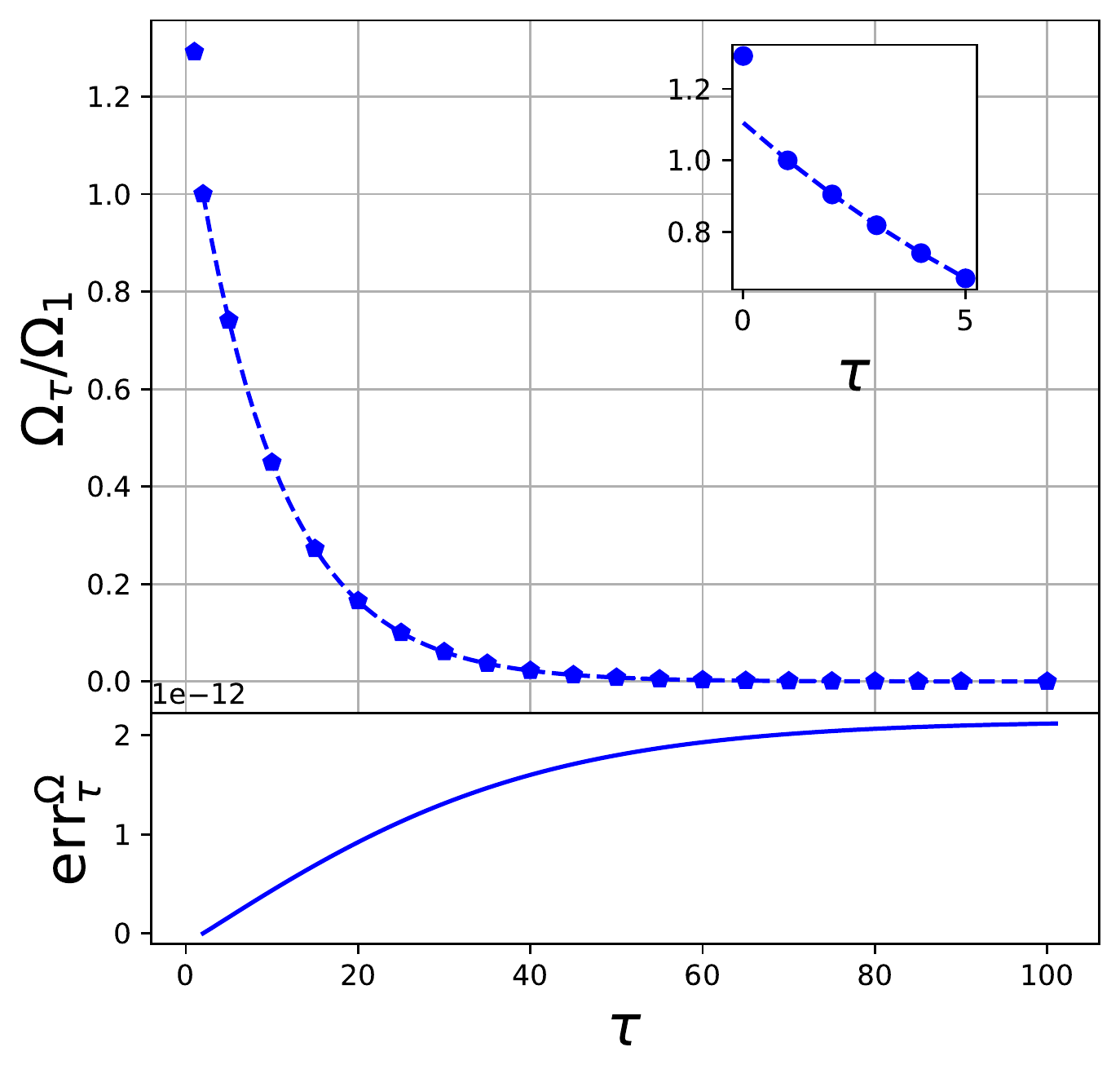}

    \caption{Numerical check of equilibrium properties for  ACFs given by $e^{-\tau/\tau_{k}}$ where $k = \{\mu,\NT\}$. We arbitrarily fixed $(\tau_\NT,\tau_{\mu}) = (10,20)$. The numerical solver has been implemented with $T_{cut} = 5\cdot 10^2$ and $T_{it} = 200$. (Left) In the upper panel we show the good collapse between  $\Xi_\tau/\Xi_0$ (bullet points) and  $ \Xi_\tau^\IT/\Xi_0^\IT$ (dashed line). The collapse between these two ACFs  is quantified  in the bottom panel, where the relative cumulative absolute error between the two curves is displayed. The inset in the top panel shows the collapse on the variogram. (Right) In the main top panel we show the good collapse for positive lags between  $ \autocvol_\tau/\autocvol_1$ (bullet points) and $ \autocvol_\tau^\NT/\autocvol_1^\NT$ (dashed line), whereas in the inset we show that the collapse doesn't involve the lag 0 term.
    In the bottom panel the collapse between these two ACFs is quantified, calculating the relative cumulative absolute error starting from lag 1. }
    \label{fig:example_mark}
\end{figure}

    The left panels of Figs.~\ref{fig:example_pl},~\ref{fig:example_osc}  and \ref{fig:example_mark} display numerical results that do confirm Eq.~\eqref{eq:price_eff}. In particular, in  top panels, bullet points correspond to $\Xi_\tau/\Xi_0$ obtained by means of the numerical solver of Eq.~\eqref{eq:master_equation} and show a good collapse on the dashed line, which corresponds to  $\Xi_\tau^\IT/\Xi_0^\IT$ calculated semi-analytically.
    In the bottom part of the panels instead we  show the relative cumulative absolute error between the two curves, defined as:
    \begin{equation}
    \label{eq:err_sigma}
    err^\Xi_\tau = \frac{\sum_{i = 0}^\tau |\Xi_i/\Xi_0 - \Xi^\IT_i/\Xi^\IT_0 |}{\sum_{i = 0}^\tau |\Xi_i/\Xi_0|}.    
    \end{equation}
    
    In Figs~\ref{fig:example_pl} and ~\ref{fig:example_osc}, where non-markovian ACFs  are examined,  these errors are larger than in Fig.~\ref{fig:example_mark}, where ACFs decay exponentially. This is due to the fact that in  the former case the forecast of future dividends suffers from finite size effects. The estimation of these effects is carried on in detail in Appendix~\ref{app:convergence}.

    The inset of the left-top panels shows the variogram of the price, defined by ${V}_\tau := \avg\left[ (p_t-p_{t+\tau})^2 \right]$, which, as expected, is linear at high frequencies and mean-reverting at low frequencies.

\paragraph{Excess demand  covariance}
 The equilibrium is characterized by an excess demand ACF  $ \autocvol_\tau  := \avg[q_t q_{t+\tau}]$  with the same temporal structure as the one related to NT's trades,   plus an extra contribution at lag 0. In formula:
    \begin{equation}
    \label{eq:camouflage} \autocvol_\tau  =a  (\tilde{\autocvol}^\NT_\tau+ \tilde{b} \delta_\tau), \ \text{with} \ \tilde{\autocvol}^\NT_0 = 1,
    \end{equation}
    where  the symbol  \(\delta_\tau\) denotes the discrete delta function, while  \(a\) and \(b\) are scalars. The excess demand  variance is given by $\autocvol_0 =a (1+\tilde{b})$.
    
    Since IT's information at time $t$ does not include the current trade of the NT  $q^\NT_t$ (see Eq.~\eqref{eq:IT_info}), the best that the IT can do in order to hide his trades is to create a trading strategy such that the excess demand ACF resembles that of the NT apart from the lag 0 term. Because of the distortion at lag 0, we call this property \emph{quasi-camouflage strategy}\footnote{Camouflage is also called inconspicuous strategy in the economics literature \cite{Inconsp,Inconsp-theorem,Taub2}}. Indeed, in order to prolong his informational advantage over the MM, the IT hides his trades in the excess demand process by creating a strategy that resembles that of the NT alone.
    
    Right panels of Figs.~\ref{fig:example_pl}, ~\ref{fig:example_osc}  and \ref{fig:example_mark} display numerical results that confirm the quasi-camouflage property. In top panels bullet points correspond to $ \autocvol_\tau/\autocvol_1 $ obtained by means of the numerical solver of Eq.~\eqref{eq:master_equation} which show a good collapse  for positive lags on the dashed line, which corresponds to $ \autocvol_\tau^\NT/\autocvol_1^\NT $. It is clear, from the insets of the plots on the left,  that the collapse is not reached at lag 0.  As it will be shown in the next section, this extra contribution at lag 0 depends in a non trivial way on the ACFs of the dividends and NT's trades. 
    On the bottom,  the relative cumulative absolute error between the two curves is presented. In this case, it starts from lag 1, so: 
    \begin{equation}
    \label{eq:err_C}
    \mathrm{err}^\autocvol_\tau = \frac{\sum_{i = 1}^\tau |\autocvol_i/\autocvol_1 - \autocvol^\NT_i/\autocvol^\NT_1 |}{\sum_{i = 1}^\tau |\autocvol_i/\autocvol_1|}.    
    \end{equation}
Again, these errors are larger in the case where non-markovian ACFs  are examined.

From the properties given by Eqs.~\eqref{eq:price_eff} and \eqref{eq:camouflage}, together with the MM's break even condition, one is in principle able to find the propagator. In fact, introducing the price ACF $\Sigma_\tau:= \avg[p_t p_{t+\tau}]$, from the definition of the propagator \eqref{eq:prop} follows that:
\begin{equation}
\label{eq:implicit_G}  \Sigma_\tau = \sum_{t'=-\infty}^{t+\tau}\sum_{t''=-\infty}^{t}G_{t+\tau-t'}G_{t-t''}\autocvol_{|t'-t''|}, \  \text{with} \ \tau>0,
\end{equation}
where the price ACF \(\Sigma_\tau\) can be computed from  Eq.~\eqref{eq:price_eff} and the excess demand ACF is given by Eq.~\eqref{eq:camouflage}.
This program can be accomplished in the case of a Markovian system and it is described in full detail in Sec.~\ref{sec:Markov}. There, we shall provide semi-analytical results for all of the parameters introduced in the equations listed above, which do share qualitative features with the general non-Markovian case. An interesting finding of this analysis is given by the fact that as the predictability of the NT's trades process increase, the IT's camouflage becomes exact, allowing him to  reduce the cost due to price impact of his trading schedule. 

But before discussing the Markovian case, let us highlight similarities and differences with respect to existing models.

\section{Relation with existing models}
\label{sec:existing_models}

\subsection{Kyle model}
While strongly inspired by the one-period Kyle model, our model is quite different on several grounds. First, instead of exogenously postulating the presence of a fundamental price, in our setting it is the integrated-dividend process  that plays the role of the fundamental price, mechanically relating it to the payoff of the asset. Second, we do not have explicit fundamental price revelation, thus allowing to consider a stationary setting in the model. Such a stationary regime is relevant in practice because in order to analyze the behavior of the market at short time scales (minutes, hours) one would like to abstract away the non-stationary effects potentially induced by the dynamics of the fundamental information (e.g, dividends, earning announcements, scheduled news) at slower time scales.
Third, we introduced (integrable) serial correlations both in the dividends~-- equivalently, in the fundamental price --~and in the order flow. 

Let us also point out how we can recover the Kyle model in our setting.
Assuming that $(i)$ the NT's trades are uncorrelated, $(ii)$ the sum of future dividends $p^\text{F}_t$ follows a random walk process, $(iii)$ the IT knows the value of $p^\text{F}_t$ at the beginning of each period and $(iv)$ $p^\text{F}_t$ becomes public information once the MM has set the price, we recover exactly an iterated version of the single period Kyle model.

\subsection{Propagator model}
Equation~\eqref{eq:implicit_G} is the cornerstone equation when dealing with propagator models. It is typically used in the literature in order to extract a propagator $G_t$ from empirical data given the order flow correlation and the price volatility. Hence, our framework allows us to recover the propagator model in an economically standard setting, with three important caveats:
\begin{itemize}
    \item The excess demand ACF  $\Omega_\tau$ function observed in real markets is typically non-integrable, due to the strongly persistent nature of the order flow~\cite{Propagator,TOTH2015218}.
    \item The price process observed in real markets is close to be diffusive at high frequency.
    \item The propagator observed in real markets is found to be a slowly decaying function of time.
\end{itemize}
Let us address these empirical facts, showing how one can account for them within our stylized model.

First, the non-integrability of the excess demand ACF can be retrieved by extending our framework to the case in which the NT's trades ACF are themselves non-integrable. This is due to the fact that the camuflage condition relating excess demand and noise trading is also expected to extend to the setting of non-integrable NT's trades.

Second, price diffusivity also can be  recovered in our model as the limiting regime in which dividends are much slower than any other time scales in the model.
In order to prove this, note that the variogram of the price can be written in terms of the price ACF $\Sigma_\tau$ as follows:
\begin{equation}
\label{eq:variogram}
 V_\tau = V_\infty (1-\tilde{\Sigma}_\tau), \ \text{where} \ \tilde{\Sigma}_0 = 1,
\end{equation}
where the first equality holds in stationary conditions, as the one described by the model introduced here.
Thus, we do recover price diffusivity at high frequency if $\tilde{\Sigma}_\tau -1 \propto \tau/\tau^*$ in the high frequency limit of the model, i.e.,  $\tau \ll \tau^*$, where $\tau^*$ is some typical timescale. Instead, in the opposite low frequency limit $\tau \gg \tau^*$, because of the assumption of mean-reverting dividends, which translates into having mean-reverting fundamental price, the price ACF decays to zero, i.e., $\Sigma_\tau \sim 0$, and we recover a flat variogram. 
For example, in the Markovian case described below, where the dividends ACF is an exponential decay function with timescale $\tau_\mu$, one has $\tau^* = \tau_\mu$.
To wrap up, if the hypothesis of linear price ACF $\Sigma_\tau$ in the high frequency limit holds, the price in our model interpolates between two very different situations: when the model is probed in its high frequency limit it describes a market with diffusive price, while in the low frequency limit the price is mean-reverting. This is very satisfactory since it is obtained with a single propagator, which is the solution of Eq.~\eqref{eq:master_equation}.
At high frequency, where the dividend process appears highly persistent, price diffusivity stems from IT's surprises in dividends variations: this is the universal mechanism which originates the diffusive behavior in our model. In fact, in this limit, the IT's estimate of the fundamental price is a martingale, and thus it is described by a diffusive process. From Eq.~\eqref{eq:price_eff} follows that the price process itself it is described by a diffusive process.

Since the first two properties can be retrieved, the third one follows from standard scaling arguments. Thus, in the high frequency limit price impact has to be a slowly decaying function of time in order to ensure price diffusion while having a strongly correlated order flow process via Eq.~\eqref{eq:implicit_G}.

It is interesting to notice that in order to observe any impact at all in the model, one is forced to introduce a non-trivial\footnote{See the brief discussion under Eq.~\eqref{eq:p_best}} dividend process: the introduction of fundamental information that gives the IT an informational advantage over the MM is enough in order to induce  non-trivial dynamics into the price, and to typically induce a diffusive behavior of prices at high frequency.
Hence, the price paid in order to micro-found the propagator model is the introduction of an auxiliary dividend process, whose detailed shape is inessential at high enough frequency, but whose fluctuations sets the scale of the price response.

\section{Markovian case}
\label{sec:Markov}
Significant simplifications of the equilibrium condition \eqref{eq:master_equation} are possible in the case in which both the dividend and the NT flow are Markovian processes, where their ACFs are given by:
\begin{subeqnarray}
\label{eq:markov}
    \Xi^\mu_\tau  &=& \Xi^\mu_{0} \alpha_\mu^\tau,
    \\
    \autocvol^\NT_\tau &= &\autocvol^\NT_0 \alpha_\NT^\tau.
\end{subeqnarray}
One of these simplifications comes from the fact that the price estimate  $p^\IT_t$ given by Eq.~\eqref{eq:p_best} is proportional to the current dividend realization, i.e.  $p^\IT_t =  \mu_{t-1} \alpha_\mu/(1-\alpha_\mu)$. Thus,  the price efficiency property given in Eq.~\eqref{eq:price_eff}  becomes:
\begin{equation}
\label{eq:price_corr_mark}
    \Xi_\tau = \Xi_0 \tilde{\Xi}^\text{F}_\tau, \ \text{with} \ \tilde\Xi^\text{F}_0 = 1,
\end{equation}
where $\Xi^\text{F}$ is the return ACF of the fundamental price $p^\text{F}_t$. From Eq.~\eqref{eq:price_corr_mark} follows that the ACF of the price process $\Sigma_\tau$ is a decaying exponential with timescale given by $\tau_\mu := -1/\log(\alpha_\mu)$. As a result, the price process in the Markovian case is a discrete Ornstein-Uhlenbeck process with timescale $\tau_\mu := -1/\log(\alpha_\mu)$.

   We validated the result of the iterative numerical solution exposed in the previous section by solving explicitly the equilibrium condition in two peculiar Markovian cases:  the case of non-correlated NT trades, obtained by replacing the equation for the NT's trades ACF by $\autocvol^\NT_\tau =\autocvol^\NT_0 \delta_\tau $, and the case in which the ACF timescale of NT's trades is the same as the dividends' one, i.e., the case given by Eq.~\eqref{eq:markov} with $\alpha_\mu = \alpha_\NT$.  These findings are reported in Appendix~\ref{app:solution_master}.

Furthermore, we found the explicit solution of the equilibrium condition by imposing the generic equilibrium properties listed in the previous section, together with the MM's break even condition given by Eq.~\eqref{eq:avg_break_even}. Details about the outcome of this procedure are given in the following sections.

Let us point out that even though the choice of  Markovian dividends and NT's trades processes is made in order to obtain analytical results and build an intuition about the system in a simple case, the main qualitative conclusions found in this section do extrapolate to  generic stationary, mean reverting processes with integrable ACFs.

\subsection{Propagator}
\label{subsec:4.1}

\paragraph{Non-correlated NT trades}
This case is particularly simple since the quasi-camouflage property given by Eq.~\eqref{eq:camouflage} becomes exact. Eq.~\eqref{eq:implicit_G} is solved by an exponential decay propagator with the same timescale as the dividends ACF, i.e., $\tau_\mu$. The amplitude of the propagator is derived in App.~\ref{app:solution_master_non_corr}.

\paragraph{Correlated NT trades}

 The solution of Eq.~\eqref{eq:master_equation} is obtained in two steps. First, we build an ansatz based on the quasi-camouflage strategy property, i.e. Eq.~\eqref{eq:camouflage} and the property about return ACF given by Eq.~\eqref{eq:price_eff}.
Details about this  are given in Appendix~\ref{app:construction_ansatz}. Then we fix the ansatz by imposing the MM's break even condition (see  Appendix~\ref{app:solution_ansatz}). The results of this procedure, described below, do match with the results of the iterative numerical solver of Eq.~\eqref{eq:master_equation}.

\begin{figure}[t!]
    \centering
    \includegraphics[width=7cm, height=6cm,scale = 0.6]{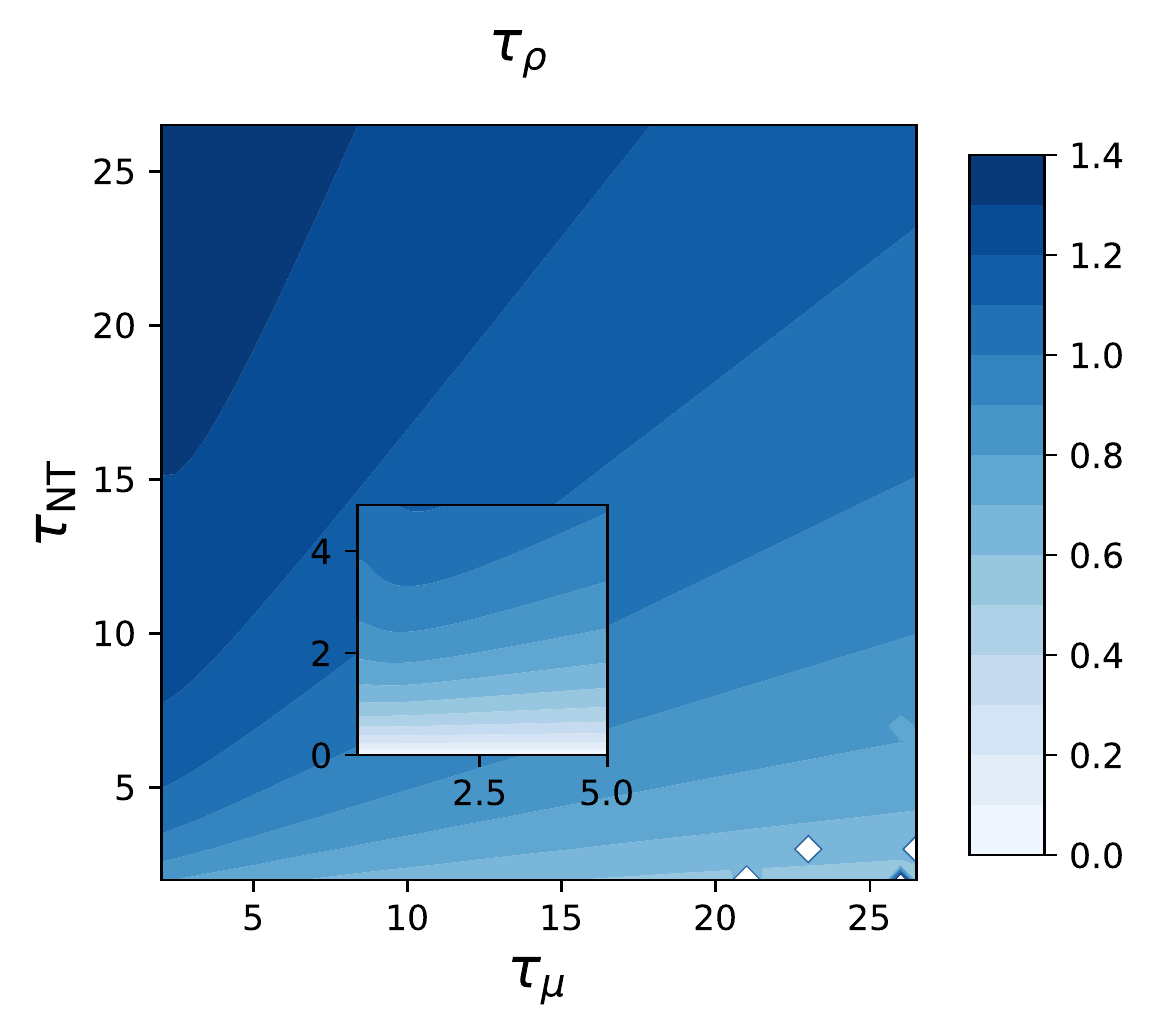}\hspace{1cm}
    \includegraphics[width=7cm, height=6cm,scale = 0.6]{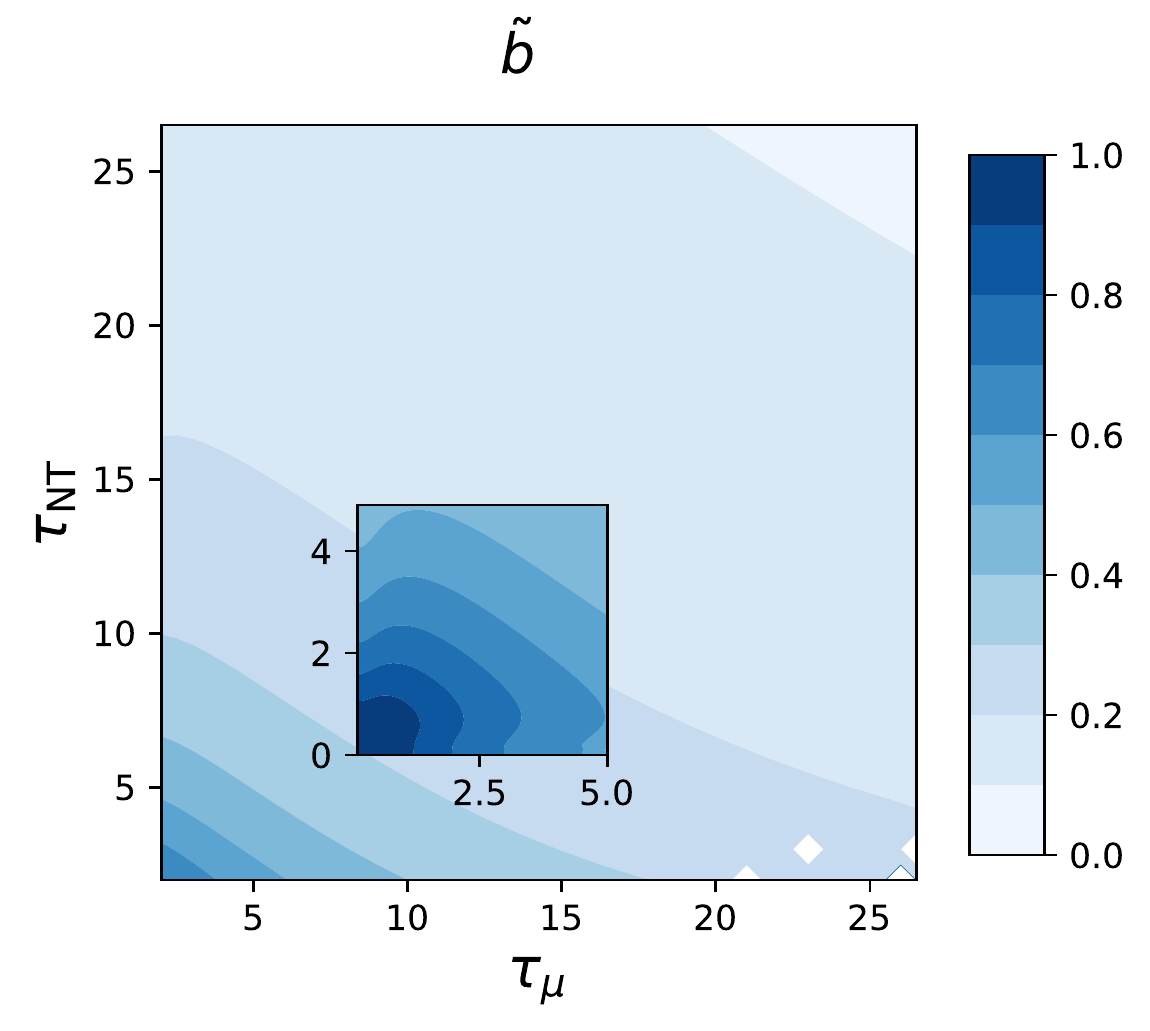}
\caption{(Left) Endogenously generated timescale  $\tau_\rho$ as a function of  \(\tau_\mu\) and \(\tau_\NT\).  $\tau_\rho$  is never larger than $\sim 2$ time-step. (Right) Amplitude of the lag 0 contribution to $\Sigma_\tau$, i.e., $\tilde{b}$ introduced in  Eq.~\eqref{eq:camouflage}, as a function of  \(\tau_\mu\) and \(\tau_\NT\). As one can see in the inset,  $\tilde{b}$ attains its maximum value for small timescale, while it decrease to zero as $\tau_\mu$ and $\tau_\NT$ increase, thus recovering exact camouflage for the IT strategy.}
    \label{fig:rho}
\end{figure}

The propagator we find   
 reads:
\begin{equation}
\label{eq:Markov_g_ansatz}
G_\tau =  G_0 \left[\frac{\alpha_\mu-\alpha_\NT}{\alpha_\mu-\rho}\alpha_\mu^{\tau} + \left(1-\frac{\alpha_\mu-\alpha_\NT}{\alpha_\mu-\rho}\right)\rho^{\tau} \right],
\end{equation}
where a new timescale \(\tau_\rho := -1/\log(\rho)\) appears. This new timescale is given, in the general Markovian case, by a non-linear combination of the two fundamental timescales \(\tau_\mu\) and \(\tau_\NT := -1/\log(\alpha_\NT)\) (the implicit  expression for $\rho$ and $G_0$ is obtained as illustrated in Appendix~\ref{app:solution_ansatz}).
From the left panel of Fig.~\ref{fig:rho}, it is clear that in the regime in which  $\tau_\mu,\tau_\NT \gg 1$, \(\tau_\rho\)  approaches a value close to the time-step, i.e. $\tau_\rho \sim 1$, thus being much smaller than the two fundamental timescales. 
This finding and the one related to the case with non-correlated NT's trades indicate that when dividends are highly persistent ($\alpha_\mu \rightarrow 1$) the propagator exhibits a quasi-permanent component and a  non-zero transient component. The former stems from the apparent persistency of the fundamental process when probed at high frequency, while the latter arises from non-trivial predictability of the NT's trades process.

As we shall see below, the large timescales behavior of $\tau_\rho$ is related to the  behavior of the excess demand ACF distortion at lag 0, i.e.  \(\tilde{b}\), introduced in Eq.~\eqref{eq:camouflage}.
 In fact, in the derivation of  Eq.~\eqref{eq:Markov_g_ansatz} (see Appendix~\ref{app:construction_ansatz}) one finds:
\begin{equation}
\label{eq:link_rho_Cq}
    \tilde{b} = \frac{\rho (1-\alpha_\NT^2)}{\alpha_\NT (1+\rho^2)-\rho(1+\alpha_\NT^2)}.
\end{equation}
In the right panel of Fig.~\ref{fig:rho} we display  \(\tilde{b}\) as  function of $\tau_\mu$ and $\tau_\NT$. This amplitude is close to 1 in the limit of small dividends and NT's trades timescales and decreases to zero as these increase. Thus, the excess demand ACF temporal structure resembles more and more the NT's one as soon as the NT's trades or dividends are strongly correlated.

The interpretation of this finding is the following: the IT wants to hide his own trades in the excess demand process, by shaping the ACF  to resemble the NT's trades one. However the IT knows only up to time \(t-1\) the realization of the NT's trades process (see Eq.~\eqref{eq:IT_info}). If this process is only weakly   correlated, the IT's information  about it does not allow a good prediction of NT's trade at time \(t\). Therefore, the IT is not able to hide his current trade. Instead, if the NT's trades are strongly correlated, the IT's information about NT's past trades allows him to accurately predict the current NT's trade, and thus the IT is able to hide his current trade. 
Briefly, we find that:
\begin{equation}
    \autocvol_\tau \rightarrow  \autocvol_0 \tilde{\autocvol}^\NT_\tau \ \text{as} \ \ \alpha_\NT \rightarrow 1,
\end{equation}
thus recovering an exact camouflage trading strategy of the IT, exhibited by many Kyle-like models \cite{Cetin, Inconsp-theorem, asymptotic_milgrom, Taub2}.

The limit \(\alpha_\NT \rightarrow 1\) and \(\alpha_\mu \rightarrow 1\) can be interpreted as the continuum limit of our discrete model. In this case, using \(\autocvol_\tau = \autocvol_0 \tilde{\autocvol}^\NT_\tau\), and  Eq.~\eqref{eq:price_corr_mark} in continuous-time one can solve the continuous-time analog of Eq.~\eqref{eq:implicit_G}, finding:
\begin{equation}
    G_\tau = G_0\left( \delta_\tau + \frac{\tau_{\mu}-\tau_\NT}{\tau_{\mu}\tau_\NT}  e^{-\tau/\tau_\mu}\right).
\end{equation}
From this equation we can see that the term in the propagator that depends on the endogenously generated timescale (see Eq.~\eqref{eq:Markov_g_ansatz}) approaches a Dirac delta function in the continuum limit of the model, as a result of the IT's exact camouflage strategy. 

\subsection{Excess demand  variance}
\label{subsec:4.2}
\begin{figure}
     \centering
     \includegraphics[width=7cm, height=6cm,scale = 0.6]{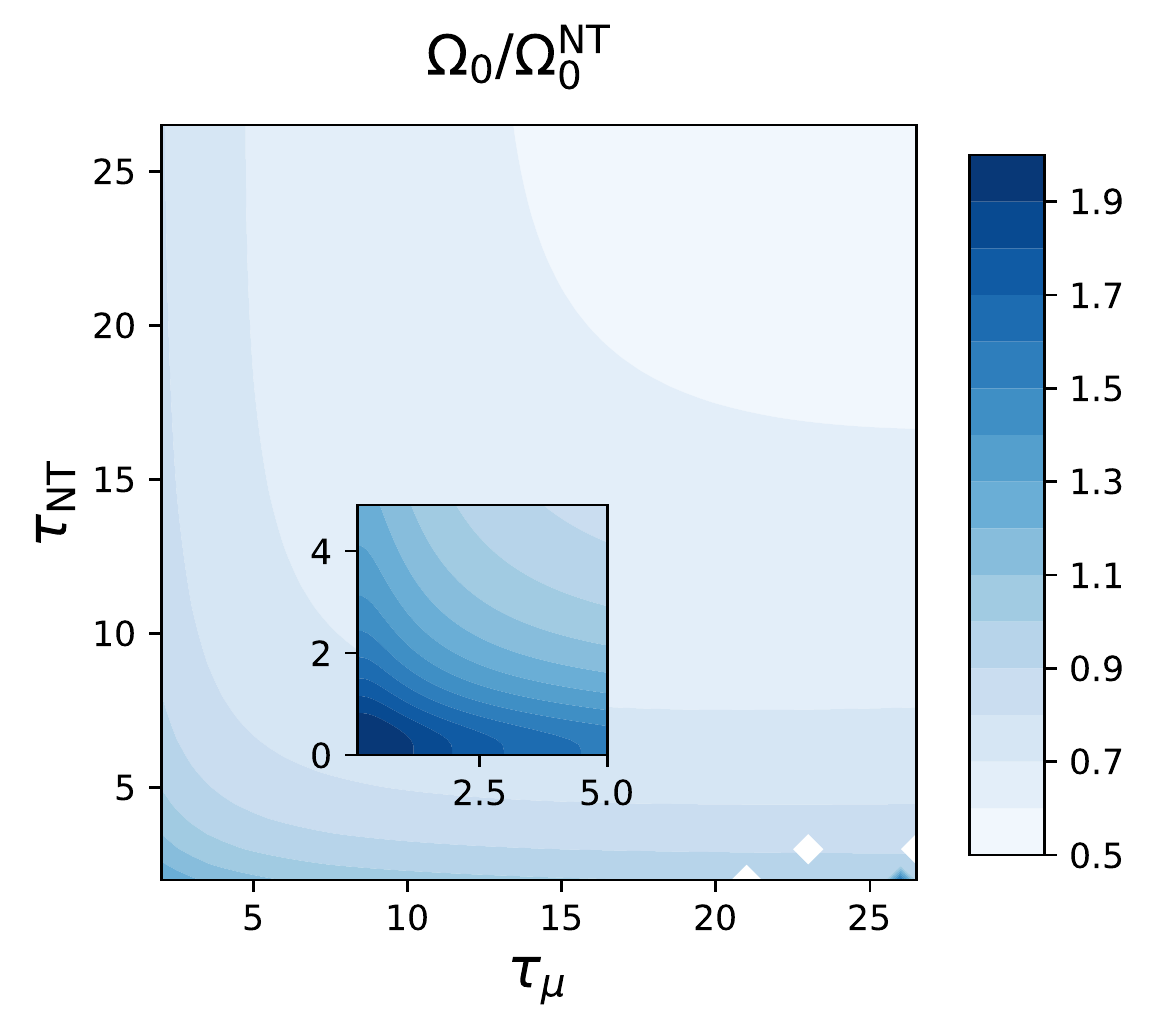}
     \hspace{1cm}
     \includegraphics[width=7cm, height=6cm,scale = 0.6]{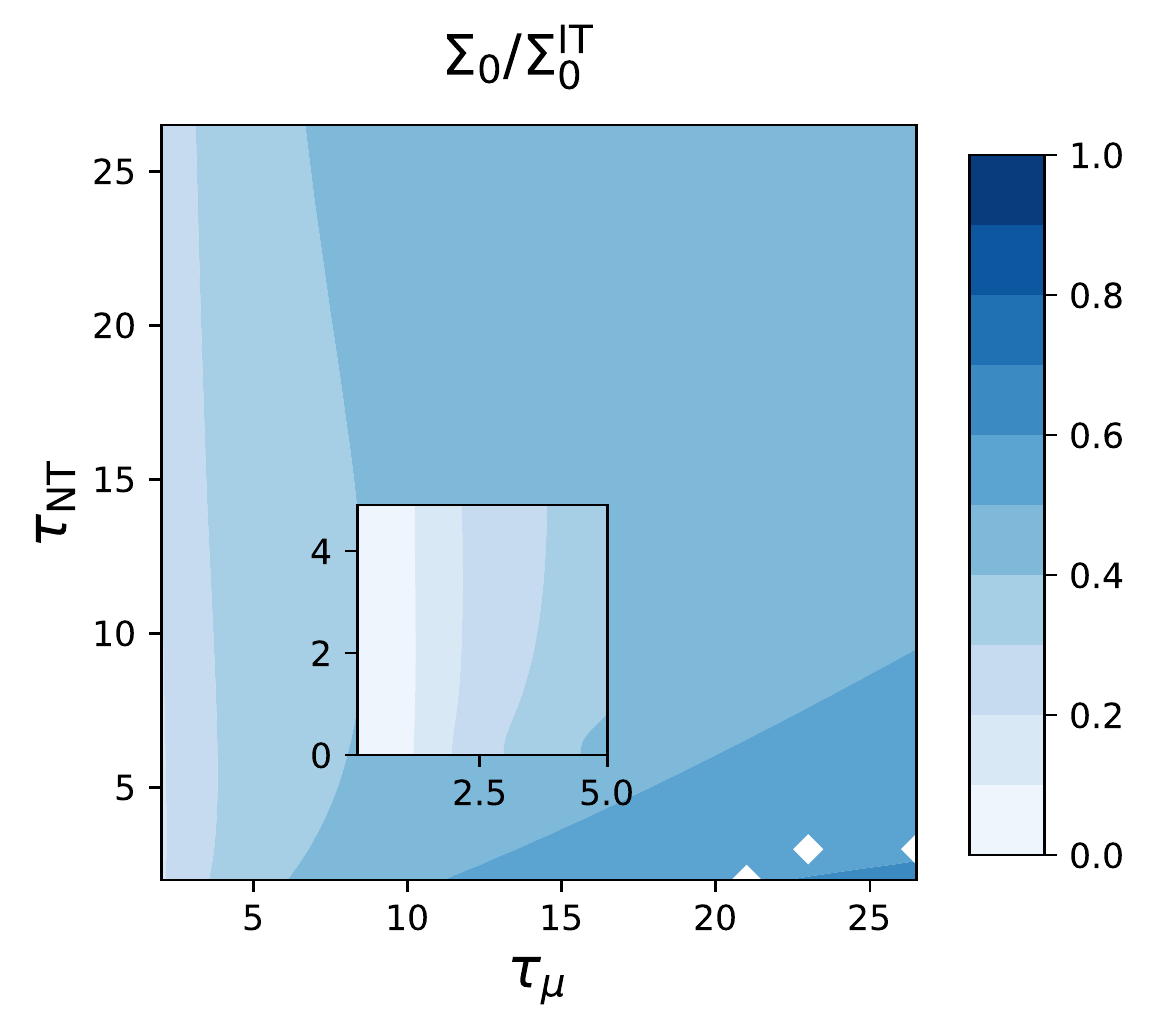}
     \caption{(Left) Ratio between variance of the excess demand and variance of the NT's order flow as a function of \(\tau_\mu\) and \(\tau_\NT\). When the timescale $\tau_\mu$ and $\tau_\NT$ are small (inset), the excess demand is higher than the one of the NT's trades alone. Conversely, when  the timescales $\tau_\mu$ and $\tau_\NT$ are large, the excess demand variance is lower than the one of the NT's alone. (Right) Ratio between the variance of the price and the variance of the IT's fundamental price estimate as a function of \(\tau_\mu\) and \(\tau_\NT\). The variance ratio in this case is very small when  $\tau_\mu$ is close to zero, while it increase as $\tau_\mu$ increases.}
     \label{fig:C0q}
 \end{figure}

The result for the ratio  \(\autocvol_0/\autocvol^\NT_0\) as a function of $\tau_\mu$ and $\tau_\NT$ is  presented  in the left panel of  Fig.~\ref{fig:C0q}.
The variance ratio is bounded between 2, for small timescales, and 0.5, for large timescales.
The increase of the ratio of variances, $\autocvol_0/\autocvol^\NT_0$,  for small $\tau_\NT$ can be understood as follows. In this regime,  the NT's current trade is almost unpredictable, thus the IT's current trade is independent of the current trade of the NT. As a consequence, the excess demand variance increases with respect to the NT's variance. 
As soon as the NT component of the order flow is predictable, the IT uses this information. 

\begin{figure}
     \centering
     \includegraphics[width=7cm, height=6cm,scale = 0.6]{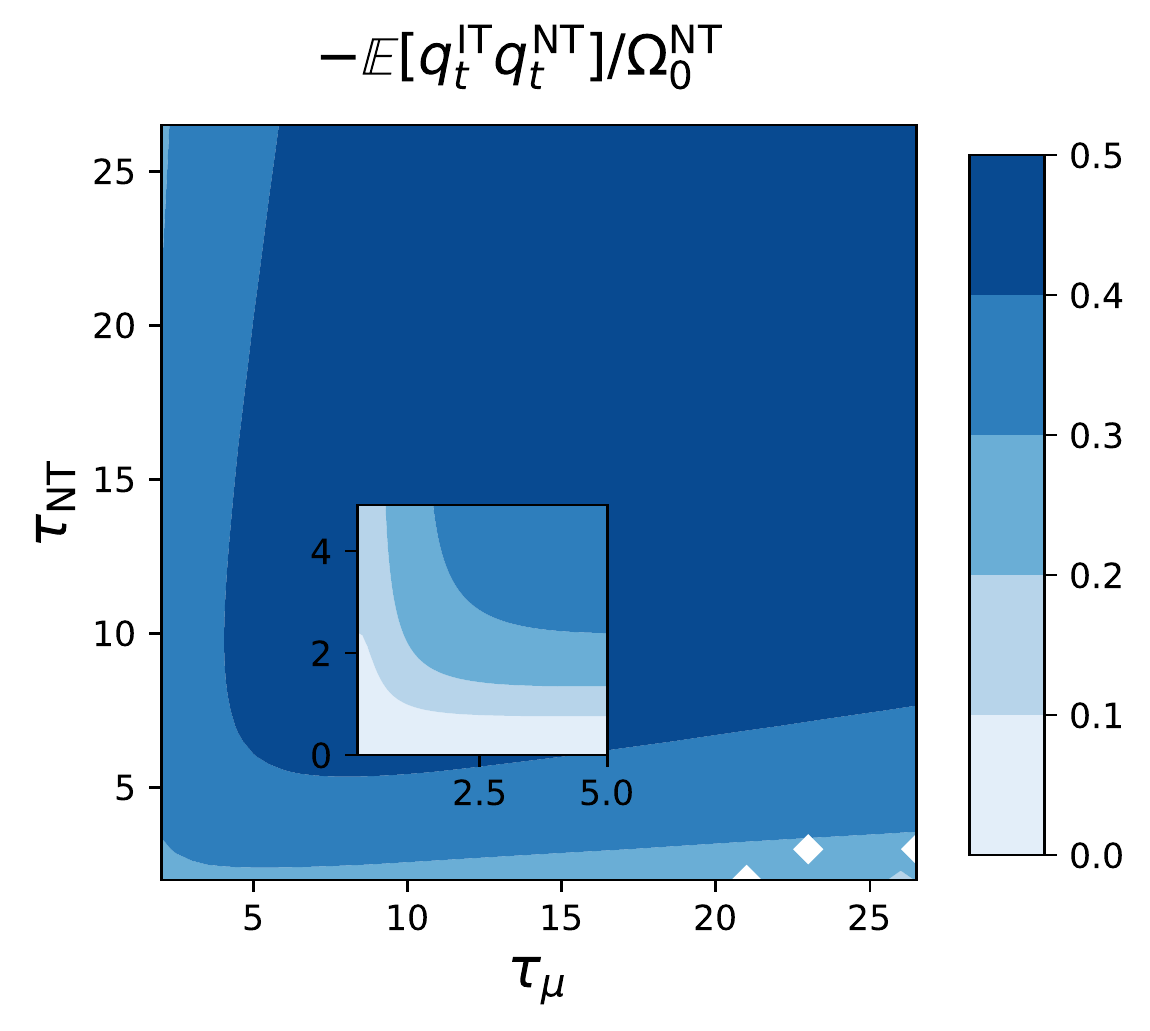}
     \hspace{1cm}
     \includegraphics[width=7cm, height=6cm,scale = 0.6]{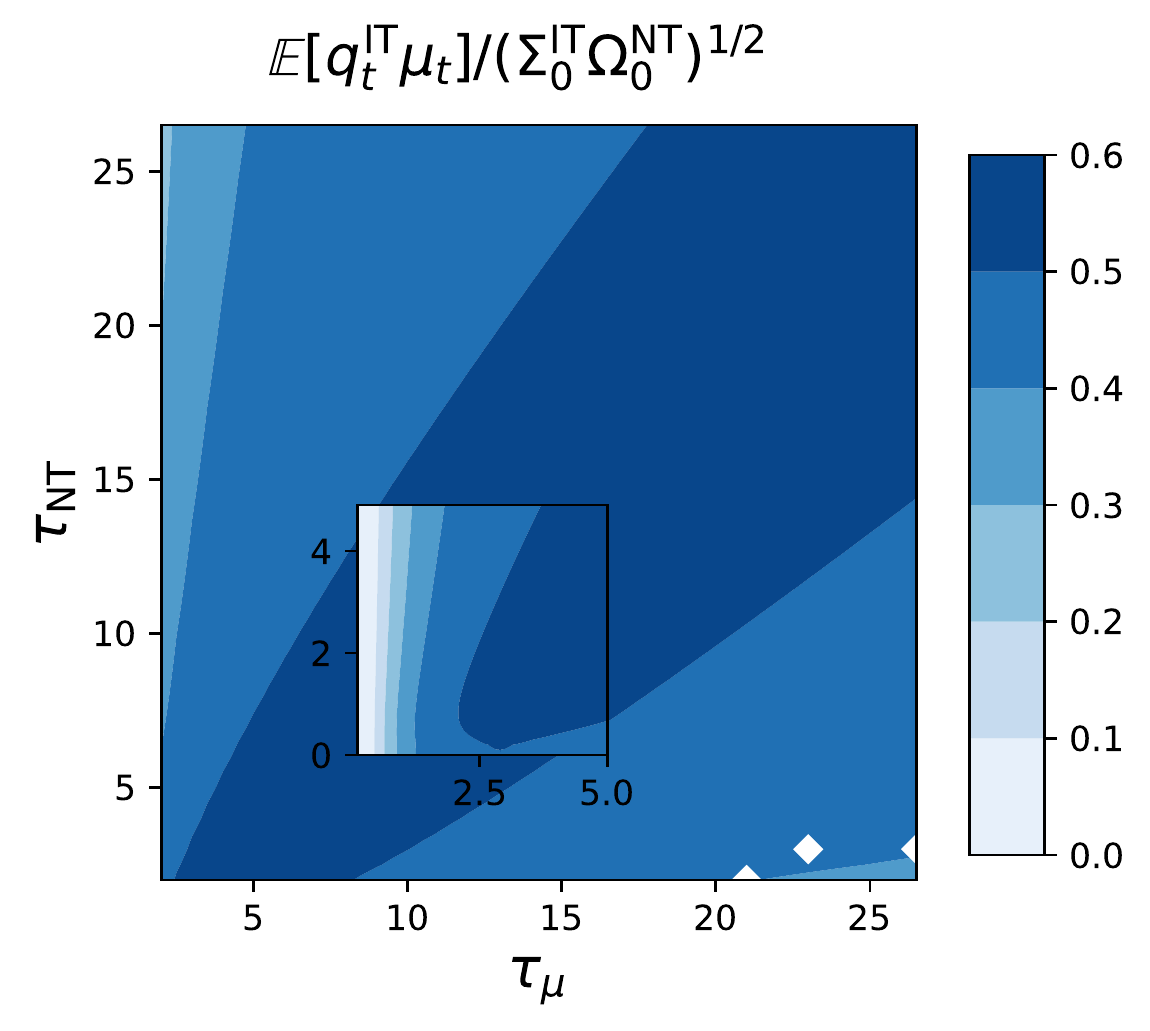}
     \caption{(Left) Ratio of the covariance between equal-time IT's and NT's  trades, and the variance of NT's trades as a function of \(\tau_\mu\) and \(\tau_\NT\). The IT's trades are anti-correlated with the equal time NT's trades. (Right) Properly rescaled  covariance of current IT's  trade and dividend  as a function of \(\tau_\mu\) and \(\tau_\NT\). The IT's trades are on average positively correlated with the equal time dividend. When the predictability of the NT's trades and dividends process increase, the current IT's trade is more positively (negatively) correlated with the current dividend (NT's trade), thus enabling him to gain more (loose less). }
     \label{fig:ITstrategy}
 \end{figure}

In particular, the IT's current trade is on average anti-correlated with the current NT trade. This enable the IT to move less the price, founding liquidity in the NT's trade and reducing the typical aggregate volume demanded to the MM. When the predictability of the NT's trades and dividends process increase, the current IT's trade is more anti-correlated with the current NT's trade, thus enabling him to loose less money due to price impact. 
The current IT's trade is instead positively correlated with the current dividend. Fig.~\ref{fig:ITstrategy} shows these findings.

\subsection{Price variance}
\label{subsec:4.3}
In our model price variance is directly linked to price efficiency, as argued below Eq.~\eqref{eq:price_eff}. As already noted by Shiller, in a Rational Expectation Model where the price is the expected fundamental price, using the principle from elementary statistics that the variance of the sum of two uncorrelated variables is the sum of their variances, one  then has $\Sigma_0/\Sigma^\text{IT}_0 \leq \Sigma_0/\Sigma^\text{F}_0 \leq 1$, where $\Sigma^\text{F}_0$ is the variance of the fundamental price.

We display the results for the  ratio $\Sigma_0/\Sigma^\IT_0$ in the right panel of  Fig.~\ref{fig:C0q}, as a function of the dividends and NT's timescales, that confirm the fundamental constraint exposed before. 
 Moreover, we find that the ratio of variances strongly depends on \(\tau_\mu\). In particular, if the dividends are weakly correlated the price variance poorly reflects the IT's price estimate variance $\Sigma^\IT_0$. Instead, in the limit of large dividend timescales with respect to the one of the NT's trades, the price variance better reflects the IT's price estimate $p^\IT_t$. In the regime of small $\tau_\NT$ and large $\tau_\mu$ the price variance accounts for all the variance of the IT's price estimate, $\Sigma^\IT_0$, as indeed found analytically from the calculations reported in Appendix~\ref{app:solution_master_non_corr}.

\subsection{Payoffs and market-making risk}
\label{subsec:4.4}
As explained around Eq.~\eqref{eq:avg_break_even}, the payoff of the different agents is, on average, the following: the MM breaks even, the NT loses and the IT gains what the NT loses. 

\begin{figure}[h!]
     \centering
     \includegraphics[width=7cm, height=6cm,scale = 0.6]{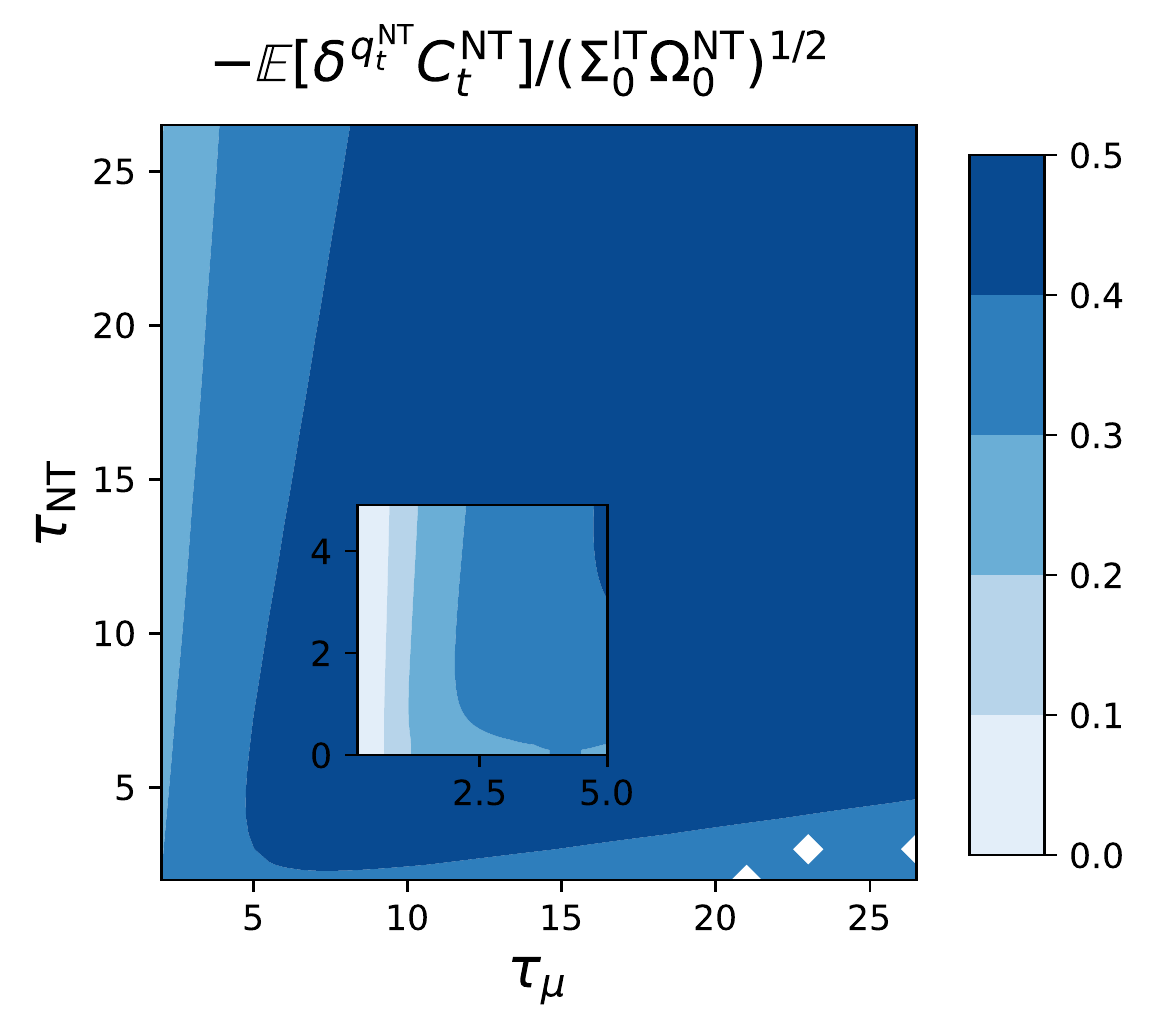}
     \hspace{1cm}
     \includegraphics[width=7cm, height=6cm,scale = 0.6]{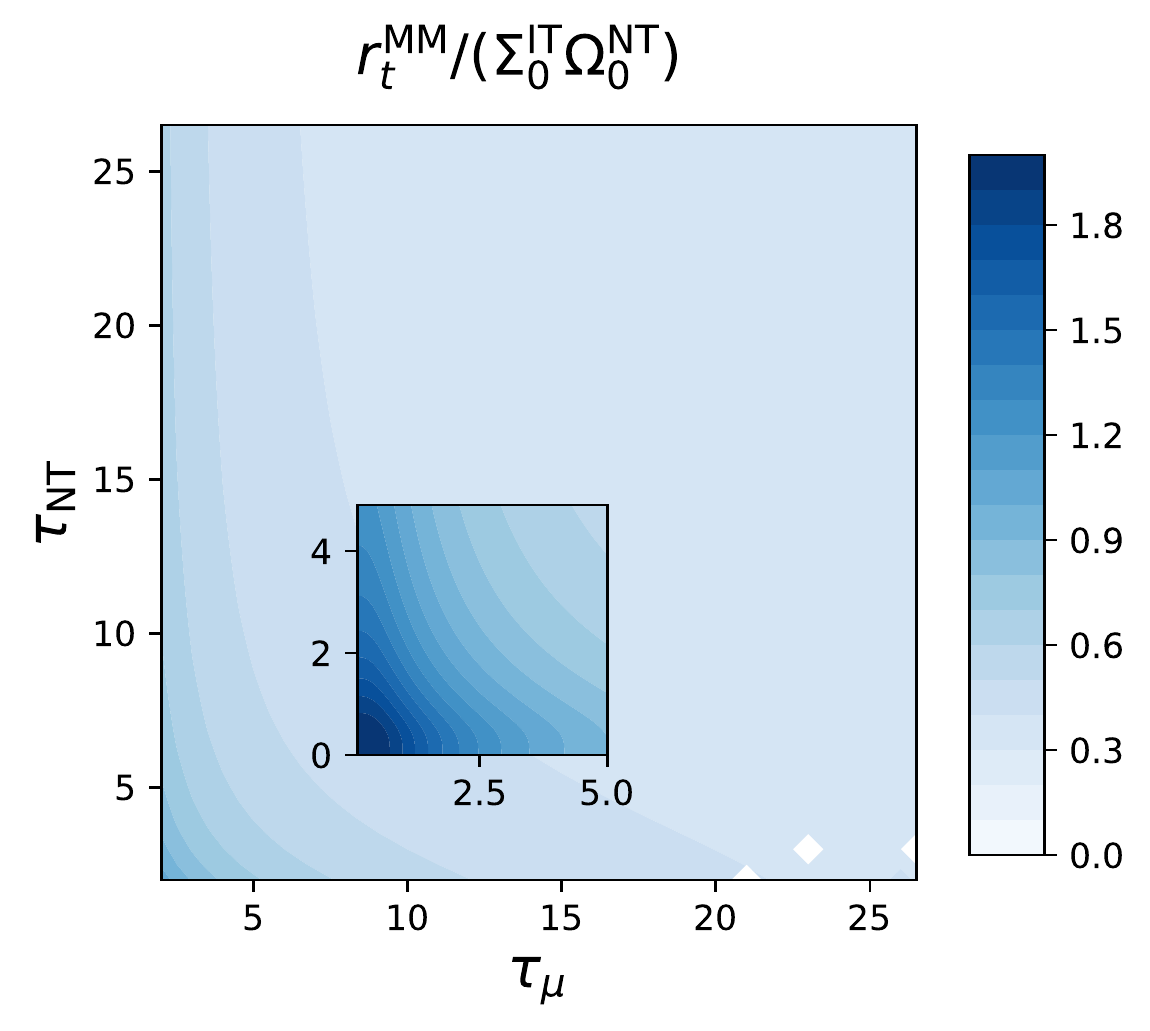}
     \caption{(Left) Properly rescaled loss per trade of the NT (or gain per trade of the IT)  as a function of \(\tau_\mu\) and \(\tau_\NT\). When $\tau_\mu$ is close to zero (inset) the loss per trade of the NT are close to zero, while these increase as the predictability of the dividends and NT's trades increase. (Right) Properly rescaled MM's risk per trade  as a function of \(\tau_\mu\) and \(\tau_\NT\). From the inset we can see that the risk is higher when the NT's trades and dividends are close to be unpredictable, whereas the risk is lower as the predictability increase.}
     \label{fig:NT_losses}
 \end{figure}

If the dividend process is completely unpredictable (but still stationary with zero-mean), then the price is set to zero by the MM; thus the IT won't trade anymore and the NT's losses are reduced to zero. 
When the $\tau_\mu$ becomes large with respect to $\tau_\NT$ (bottom right corner of main left panel of Fig.~\ref{fig:NT_losses}),  the price is more and more efficient as we have seen in the previous section. In this case,   the IT's gains are lowered, as well as the NT's losses. These findings are reported in the left panel of Fig.~\ref{fig:NT_losses}, where we plot the ratio $-\avg[\delta^{q^\NT_t}C^\NT_t]/(\Xi^\mu_0\autocvol^\NT_0)^{1/2}$, with $\delta^{q^\NT_t}C^\NT_t = -q^\NT_t \left( p_t-\sum_{t'\geq t} \mu_{t'} \right) $.
 
 Another interesting  quantity is  the risk per trade experienced by the MM, i.e.  $r^\MM_t = \avg[(\delta^{q_t}C^\MM_t)^2]$, where $\delta^{q_t}C^\MM_t = q_t \left(p_t-\sum_{t'\geq t} \mu_{t'} \right)$.
 We find:
\begin{equation}
    \label{eq:riskMM}
    r^\MM_t =  \avg[q_t^2]          \ \avg\left[\left(p_t-p^\IT_t\right)^2\right],
\end{equation}
 where we used the break even condition (Eq.~\eqref{eq:avg_break_even}) together with Wick's theorem to calculate higher order correlations of a Gaussian process.
The analytical solution is given in the right panel of Fig.~\ref{fig:NT_losses}.
As we can see, the risk experienced by the MM is high when both the timescales of the two fundamental processes are small, while it decreases when both the dividends and the NT's trades becomes predictable.

\section{Conclusion}
\label{sec:conclusions}

The aim of this paper was to provide an economically standard micro-foundation for linear price impact models, customarily used in the econophysics literature.
To do so, we presented a multi-period Information-Based Model and we analyzed its equilibrium. The model is built by generalizing the seminal Kyle model, which constitutes a theoretical cornerstone of market microstructure. First, we removed the assumption of  fundamental price revelation, assuming that a stock pays dividends to the owner but only the insider collects and exploits information about past dividends. Then, we modeled the dividends process and the noise trader trading schedule as stationary stochastic processes. In order to regularise the model we assumed that the dividend ACF  was integrable,  to ensure a bounded fundamental price of the traded stock. The model appeared to exhibit a stationary equilibrium, which we have investigated in detail. 
A self-consistent equation for the pricing-rule set by the market-maker has been derived and solved numerically. 
Two robust properties have been found: the price ACF retains the same temporal structure as the insider's fundamental price estimate and the insider strategy respects a quasi-camouflage condition, i.e., the ACF of the excess demand retains the temporal structure of the noise trader's one apart from the lag 0 term. 

As a consequence of these findings, we have been able to establish a precise correspondance between the propagator model and the Kyle model: the propagator model arises here as the high-frequency limit of a suitably stationarized Kyle model. The price impact function that is found in this regime displays a quasi-permanent component related to the timescale of variation of the fundamental information, and a transient one whose timescale is set by the persistence of the order flow.

 The assumption of stationary dividends with integrable ACF translates into having a mean-reverting price process. Since price diffusivity can be retrieved in the high frequency limit, the model is able to provide a stylized picture of what happens in real markets at high and low frequency.  The model alludes also to a relation between the  diffusion constant of the price process and the timescale over which the fundamental price mean reverts. We leave the empirical check of this finding as an interesting follow-up of the present investigation. 

The minimal model exposed here can be extended in several different ways.

First, further elements of realism (risk-aversion, spreads) could be integrated progressively, in order to see how much our main qualitative findings are impacted these effects. Similarly, the non-linear, concave nature of impact \cite{Donier} should be reconciled with our stylized, linear vision of the market.

The passive agent can be promoted to a rational agent that tracks a given target portfolio, introducing an element consistent with actual actions in real-world
markets, which could itself create long-range correlated order flow. 

Dividend revelation in real markets is infrequent, so another extension of the model proposed here is to take explicitly this fact into account, similarly to what is done in Ref.~\cite{Infrequent_auctions}, where market closure is explicitly taken into account. 

Finally, in order for our model to be able to account for the excess volatility puzzle, we need to relax the assumption on either rationality or information used by the agents that populate our universe. This can be done in different ways: for example, as in Ref.~\cite{Timmermann}, we can relax the assumption of perfect structural knowledge: for example, we can assume that the agents do not know all the parameters that define the dividend process and they try to infer these starting from actual observation. Another interesting path would be, along the line of Ref.~\cite{Hommes}, to assume that the agents decides their demand according to a misspecified equation of motion for the price. We look forward to exploring some of these interesting topics in the near future.

\section{Acknowledgments}
\label{sec:ack}

We thank J.-P. Bouchaud and C.-A. Lehalle for fruitful discussions. This research was conducted within the Econophysics \& Complex Systems Research Chair, under the aegis of the Fondation du Risque, the Fondation de l’Ecole polytechnique, the Ecole polytechnique and Capital Fund Management. 

\clearpage

\bibliography{references}

\clearpage

\appendix

\section{Numerical solver}
\label{sec:numerical_solver}
The iterative numerical scheme is as follows: 
\begin{enumerate}
\item Choose a maximum time $T_{cut}-1$ which is the maximum time-lag at which the propagator can be evaluated. In doing so the propagator is a vector of $T_{cut}$ elements. 
\item Choose a ``seed'' propagator. 
\item Plug this seed in the r.h.s. of Eq.~\eqref{eq:master_equation}.
\item Insert the result obtained with this procedure in the r.h.s. for a number of iterations equal to $T_{it}$, checking for convergence.
\end{enumerate}

The only issue of this procedure is the following: as one can see from the first of Eqs.~\eqref{eq:IT_response}, in order to compute $\vecdmd_\past$ one has to evaluate the block matrix given by $\matprop_{\fut,\past}$. This matrix has entries that cannot be calculated, due to the truncation constraint of our numerical procedure. Nevertheless, because of the mean reverting assumption of the dividends, we know that the propagator should decay to zero at large times, so the large lags terms in $\matprop_{\fut,\past}$ can be simply set to zero.
\subsection*{Convergence}
\label{app:convergence}

In this section we give further details about the convergence of the results of the iterative numerical solution of Eq.~\eqref{eq:myopic}.

\begin{figure}[h!]
    \centering
    \includegraphics[scale = 0.45]{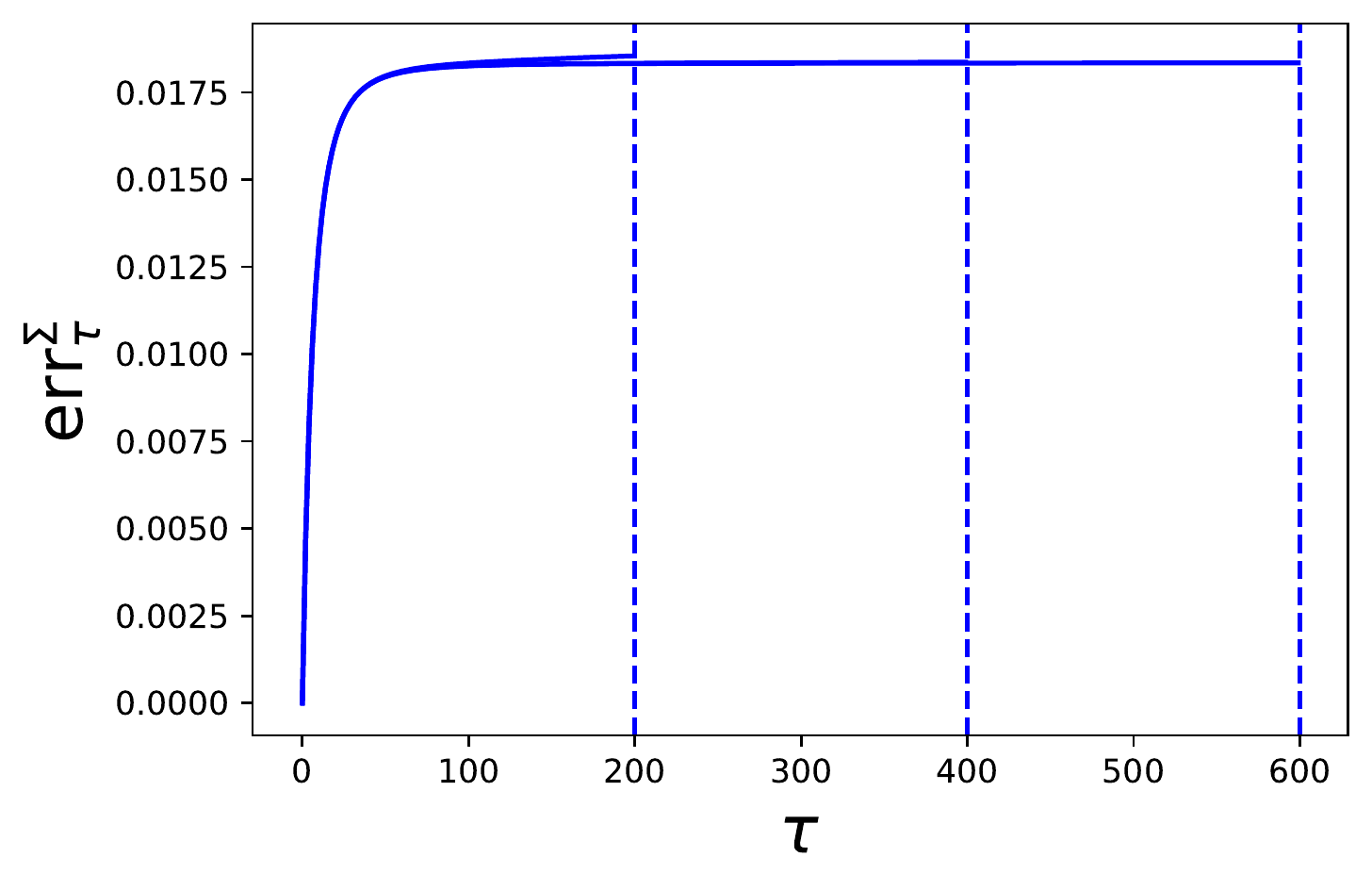}
    \hspace{1cm}
    \includegraphics[scale = 0.45]{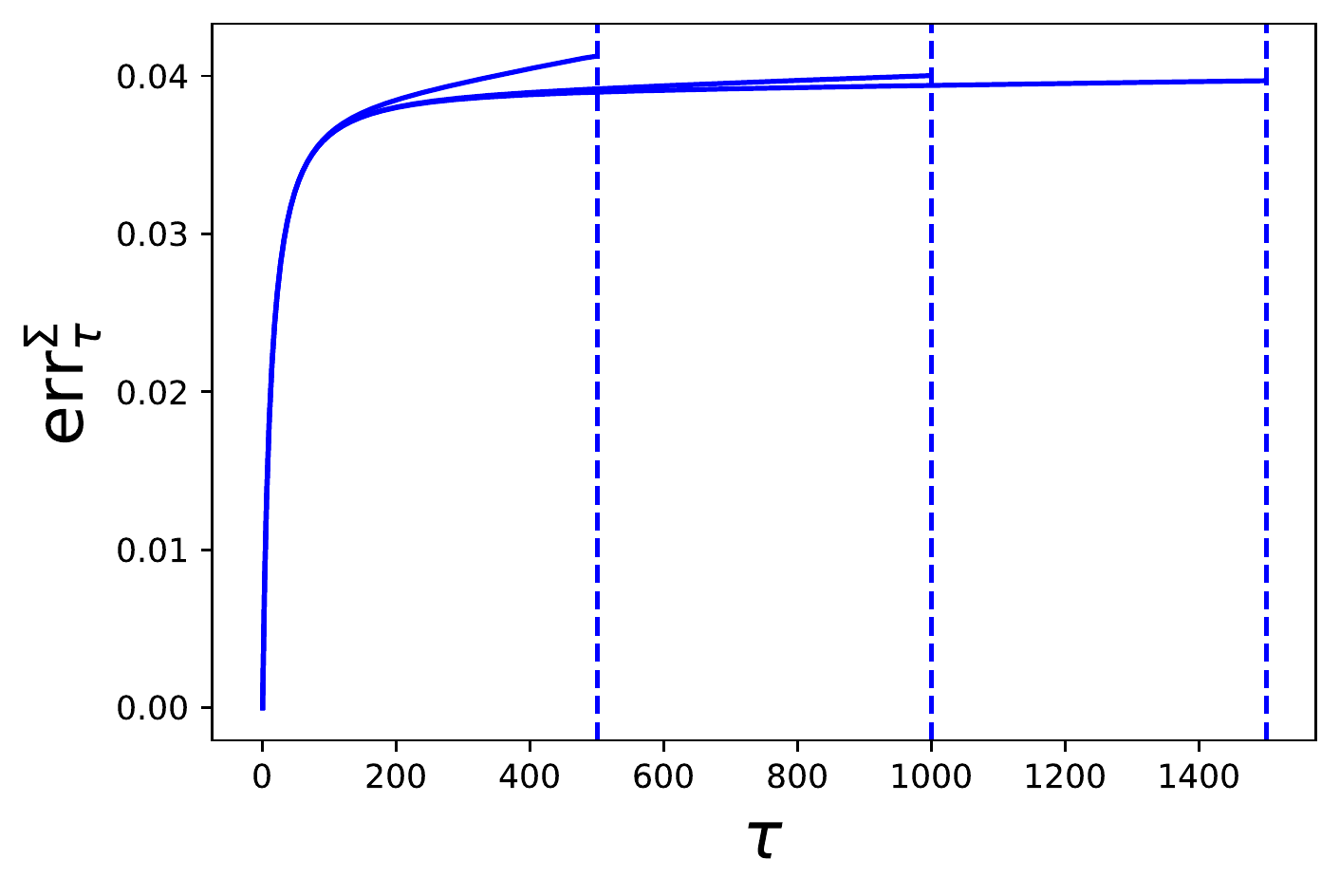}

    \includegraphics[scale = 0.45]{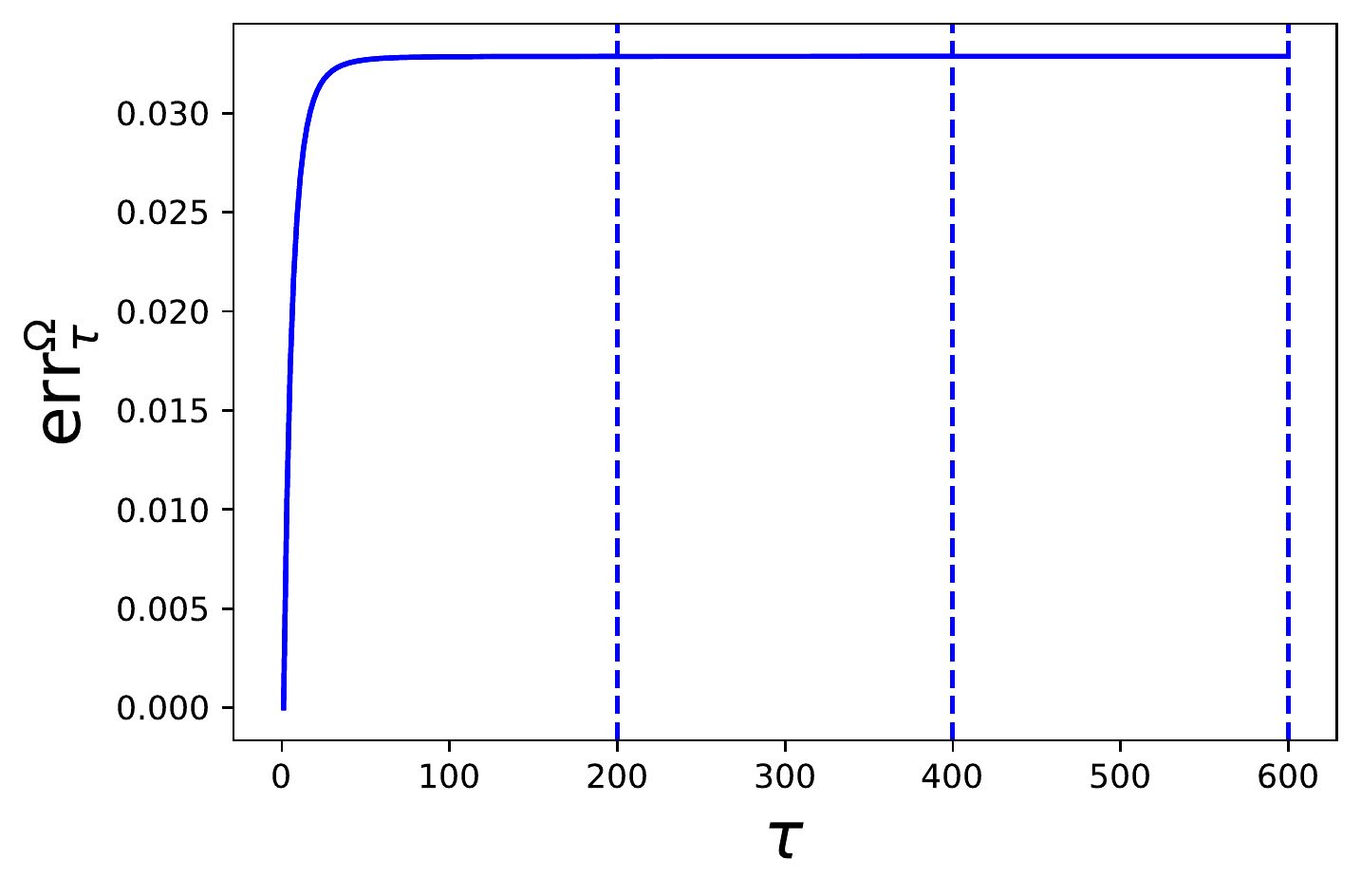}
    \hspace{1cm}
    \includegraphics[scale = 0.45]{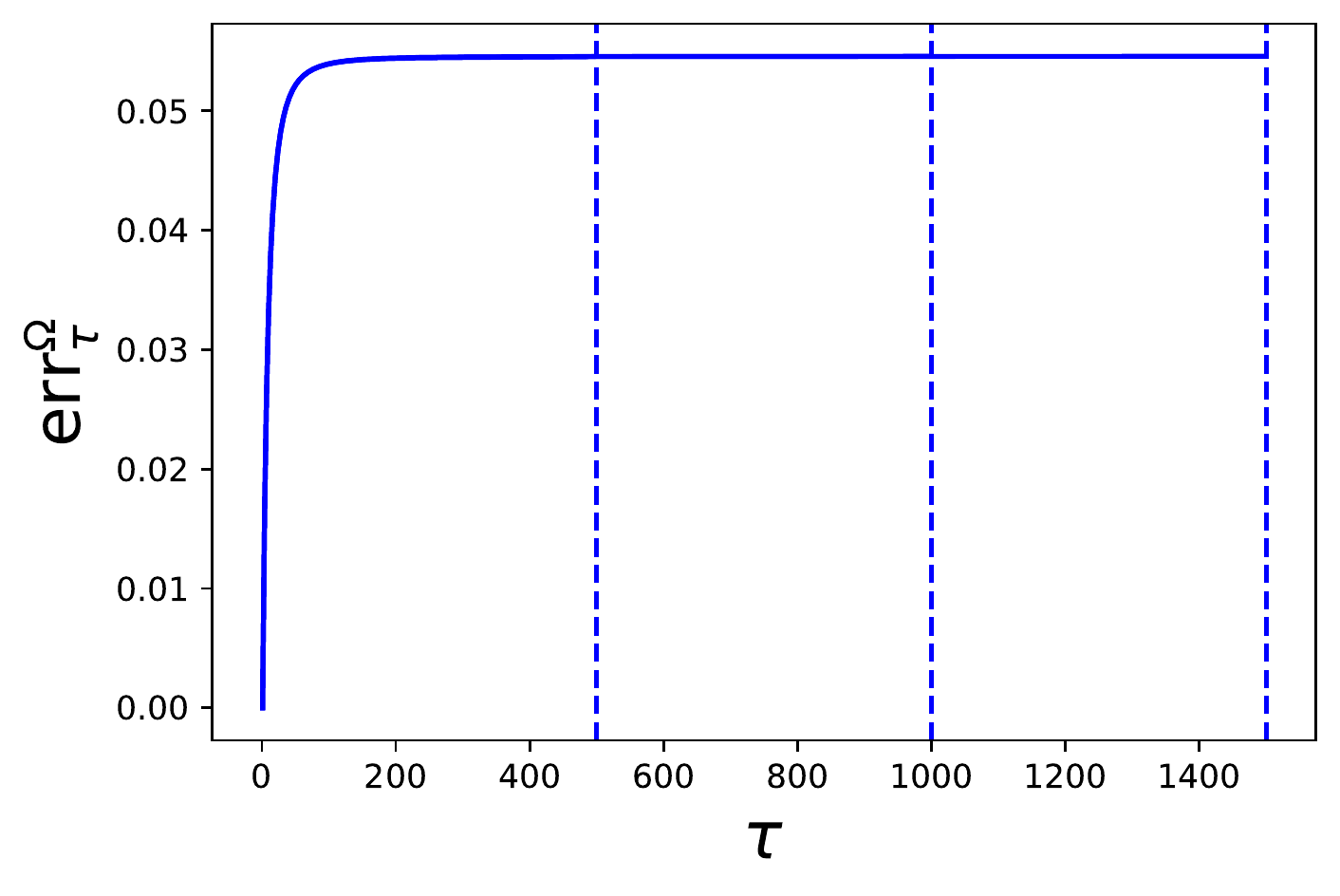}

    \caption{Numerical check of equilibrium properties with  ACFs given by $(1+|\tau|/\tau_{k})^{-\gamma_k}$ where $k = \{\mu,\NT\}$. We arbitrarily choose $\tau_\NT = \tau_{\mu} = 10$. (Left panels) $ \gamma_\NT = \gamma_\mu = 5$ and $\Delta t = 200$. (Right panels) $ \gamma_\NT = \gamma_\mu = 3.5$ and $\Delta t = 500$.}
    \label{fig:convergence}
\end{figure}

In Fig.~\ref{fig:convergence} we show  results about the relative cumulative absolute error for the price ACF $\Sigma_\tau$ and the excess demand ACF $\Omega_\tau$. The first one is calculated as in Eq.~\eqref{eq:err_sigma}, while the second one is given by \eqref{eq:err_C}.  

We choose $T_{it} = 100$, power law ACFs   for dividends and NT's trades. We plot the result  for $T_{cut,i} = \Delta t\times i$, for different $i$. 
The plots on the left are obtained with a power law  ACF that decays faster than the one used to obtain the plots on the right. We can see, as expected,  that the slower is the decay of the power law, the slower is the convergence.

We have investigated the behavior of the error for higher $T_{it}$, but we didn't find quantitative differences.

\section{Particular solutions of equilibrium condition in the Markovian case}
\label{app:solution_master}
\subsection{The case of non-correlated Noise}
\label{app:solution_master_non_corr}

In the case of non correlated NT's trades, the IT's forecast of future NT's trades is zero, and so the demand kernel $\matdmd^\NT$, explicitly given in  Eqs.~\eqref{eq:R_NT}, is zero. Since we are dealing with a Markovian dividend process, the IT's forecast at time $t$ of future dividends relies only on the last known dividend, i.e., $\mu_{t-1}$ and so $\mathsf{R}^\mu = R^{\mu} \id$, where $R^{\mu}$ is a scalar.

The self-consistent equilibrium condition given by Eq.~\eqref{eq:master_equation} for the dimensionless propagator is given by:
\begin{equation}
\label{eq:master_equal_timescales}
\tilde{G}_{t-t'} = \frac{1}{ 1-\alpha_\mu} \mathbf{e}^\top_t \tilde{\mathsf{\Gamma}} (\id-\matdmd \matL),  
\end{equation}
where
\begin{equation}
\label{eq:Gamma}
    \tilde{\mathsf{\Gamma}} = \left[ (\tilde{\mathsf{\Xi^\mu}})^{-1} + (\tilde{\matdmd}^\mu \matL)^\top \tilde{\matdmd}^\mu \matL \right]^{-1} (\tilde{\matdmd}^\mu \matL)^\top.
\end{equation}
The solution of Eq.~\eqref{eq:master_equal_timescales} is constructed in three steps. i) First we analyze the vector $\mathbf{e}^\top_t \tilde{\mathsf{\Gamma}}$ and we show  that it is related to the inverse of a tri-diagonal matrix with modified corner elements, for which the explicit expression is known~\cite{tridiagonal_inverse}. Then, ii) we prove that a single  exponential propagator solves Eq.~\eqref{eq:master_equal_timescales} and we identify  the amplitude and the timescale of the propagator  in terms of $\alpha_\mu$ and  $R^\mu$. iii) Finally, we can calculate the expression of \(R^\mu\) in terms of $\alpha_\mu$ from its general expression given in Eqs.~\eqref{eq:IT_response}. In this way we fix  completely the shape of the propagator only in terms of \(\alpha_\mu\).

i)  Since in the Markovian case \(\mathsf{R}^\mu\) is proportional to the identity matrix, from Eq.~\eqref{eq:Gamma} we obtain:
\begin{equation}
\label{eq:  app_b}
\mathbf{e}^\top_t \tilde{\mathsf{\Gamma}}  = (a_t,\mathbf{b}_{t-1}) (\tilde{\matdmd}^\mu \matL)^\top = \tilde{R}^\mu \mathbf{b}_{t-1},
\end{equation}
where the vector \(\mathbf{b}_{t-1}\) can be found by means of the block matrix inverse formula applied to the matrix inside the square brackets of Eq.~\eqref{eq:Gamma}, given by:
\begin{equation}
\label{eq:block}
  M = (\tilde{\mathsf{\Xi^\mu}} )^{-1} + (\tilde{\matdmd}^\mu \matL)^\top \tilde{\matdmd}^\mu \matL = 
\left[
\begin{array}{c|c}
a_t & \mathbf{B}_{t-1} \\
\hline
\mathbf{B}_{t-1}^\top & \mathsf{C}
\end{array}
\right].
\end{equation}
In particular, using the block inverse formula, the vector \(\mathbf{b}_{t-1}\) is  given by
\begin{equation}
\label{eq:b}
\mathbf{b}_{t-1} = -a^{-1}_t\mathbf{B}_{t-1}^\top (M/a_t)^{-1} = \alpha_\mu \mathbf{e}^\top_{t-1} (M/a_t)^{-1},
\end{equation}
where the last equality has been obtained with the following property (checked by direct inspection of Eq.~\eqref{eq:block}): $\mathbf{B}^\top_{t-1} \propto \mathbf{e}^\top_{t-1}$.   $(M/a_t)$ is the Schur's complement of $M$ with respect to $a_t$, which is given by
\begin{equation}
(M/a_t) = \mathsf{C}-\mathbf{B}_{t-1}^\top a_t^{-1}\mathbf{B}_{t-1} = (\tilde{\mathsf{\Xi^\mu}})^{-1}+(\tilde{R}^\mu)^2\id.
\end{equation}
$(M/a_t)$ is a tri-diagonal matrix with modified corner elements. 
Thus, the inverse of the Schur's complement of M with respect to $a_t$ can be calculated explicitly (see Ref.\cite{tridiagonal_inverse}).
 The explicit expression of Eq.~\eqref{eq:b} is given by a single decaying exponential: 
\begin{equation}
\label{app:gamma}
\mathbf{b}_{t-1} = b_0 \{\gamma^\tau\}_{\tau= 0}^\infty, \end{equation}
where   
\begin{equation}
    b_0 = \alpha_\mu \frac{(\tilde{R}^\mu)^2-g}{(\tilde{R}^\mu)^4},  \ \  \ \gamma = \frac{g \alpha_\mu}{(\tilde{R}^\mu)^2}
\end{equation}
and $g$ is given by:
\begin{equation}
\label{c1}
g = \frac{\beta-\sqrt{\beta^2-4}}{2 (\tilde{R}^\mu)^{-2} \alpha_\mu}, \ \ \ \beta = \frac{ (\tilde{R}^\mu)^{-2}+ 1 +(\tilde{R}^\mu)^{-2}\alpha_\mu^2-\alpha_\mu^2}{(\tilde{R}^\mu)^{-2} \alpha_\mu},
\end{equation}
so that  \(\mathbf{b}_{t-1}\) is completely specified by \(\alpha_\mu\) and \(\tilde{R}^\mu\).

ii)
We are going to proof that an ansatz for the propagator given by a decaying exponential towards zero actually solves Eq.~\eqref{eq:master_equal_timescales}. The ansatz for the propagator reads: 
\begin{equation}
\label{eq:G_exp}
G_{t-t'} = G_0 \rho^{t-t'}.
\end{equation}
As a preliminary result, 
from this ansatz, one can compute the elements of the vector $\vecdmd_t$, which appear in Eq.~\eqref{eq:master_equal_timescales},  by means of the first equation in Eqs.~\eqref{eq:IT_response} . This is given by:
\begin{equation}
R_{t-t'} =  - R_0 \rho ^{t-t'}, 
\end{equation}
where
\begin{equation}
\label{g1}
R_0 = 1-g_s, \ \ g_s = \frac{1-\sqrt{1-\rho^2}}{\rho^2}.
\end{equation}

Equipped with this result, together with Eq.~\eqref{app:gamma} one can easily show that Eq.~\eqref{eq:master_equal_timescales} is solved with the ansatz given by Eq.~\eqref{eq:G_exp}. The ansatz is constraint to satisfy the following equations:
\begin{equation}
\label{G0}
\tilde{G}_0 = \frac{b_0}{1-\alpha_\mu}, \ \rho = \frac{\gamma}{(1-R_0)}.
\end{equation}

iii) Since we proved that \(G\) is of the exponential form we are now able to compute the explicit form of \(\matdmd_\mu\), starting from its definition in Eq.~\eqref{eq:IT_response}. The explicit expression for $\tilde{R}_\mu$, which completely specifies $\matdmd^\mu$,  is given by:
\begin{equation}
R^\mu = \frac{1}{(1-\alpha)G_0} \left( \alpha_\mu g_s -\frac{\alpha_\mu^2}{\rho} \frac{1-(2-\rho^2)g_s}{1-g_s\rho \alpha_\mu} \right)
\end{equation}
Now, we can use insert in the above equation the expression for \(g\) , \(g_s\), \(G_0\), \(\rho\) given respectively by Eqs.~\eqref{c1}, \eqref{g1} and \eqref{G0} and solve for $\tilde{R}^\mu$.
In doing so we find 
\begin{equation}
\tilde{R}^\mu = 1.
\end{equation}

Finally, reintroducing the variance terms, i.e., using  \(\autocvol_\tau^{\NT}= \autocvol_0^\NT\delta_{\tau}\) and $\Xi^\mu_\tau = \Xi^\mu_0 \alpha_\mu^\tau$, then the solution to Eq.~\eqref{eq:master_equal_timescales} is given by
\begin{equation}
\label{eq:prop_NT_non_corr}
    G_{\tau} = \left( \frac{\Xi^\mu_0}{\autocvol_0^\NT} \right)^{1/2}\frac{\alpha_\mu}{1-\alpha_\mu}\left( 1-\frac{1-\sqrt{1-\alpha_\mu^2}}{\alpha_\mu^2} \right) \alpha_\mu^\tau.
\end{equation}

\subsection{The case of Noise and Signal with equal autocovariance timescales}
\label{app:solution_master_equal_timescales}

In this section we deal with the Markovian case specified by Eqs.~\eqref{eq:markov} with $\alpha_\mu = \alpha_\NT$.  A difference with the previous case is given by the fact that now $\matdmd^\NT = R^\NT\id$, where $R^\NT$  is a nonzero scalar. 
The solution of the self-consistent equilibrium condition \eqref{eq:master_equation} is akin to the one exposed in the previous section, due to a simplification induced by the assumption given by $\alpha_\mu = \alpha_\NT$. 
In order to show this we define $\mathsf{E}^\NT$ as:
\begin{equation}
    \mathsf{E}^\NT_t = \left(\id+\matdmd^\NT \matL \right) \mathsf{\autocvol}^\NT \left(\id+\matdmd^\NT \matL \right)^\top.
\end{equation}

The simplification is the following:
\begin{equation}
\begin{split}
    \left\{\left[ (\mathsf{\Xi^\mu})^{-1} + (\matdmd^\mu \matL)^\top (\mathsf{E}^\NT)^{-1} \mathsf{\matdmd}^\mu \matL \right]^{-1} (\matdmd^\mu \matL)^\top(\mathsf{E}^\NT)^{-1}\right\}_{t,t'} &=
    \alpha_\mu \matdmd^{\mu} \matL \left\{\left[ \mathsf{E}
    ^\NT(\mathsf{\Xi^\mu})^{-1} + (R^{\mu})^2 \id \right]^{-1}\right\}_{t,t'},
\end{split}
\end{equation}
where the matrix inside the square bracket on the r.h.s. is a tri-diagonal matrix with modified corner elements, for which, as seen before, analytical results are available. Thus, akin to the previous case, a propagator given by a single exponential decay term given by Eq.~\eqref{eq:G_exp} is a solution of the self-consistent equation for the propagator \eqref{eq:master_equation}.
The result of the calculation that we do not report here is given by
\begin{equation}
    R^\NT: (R^\NT)^4-3 (R^\NT)^2
   \alpha_\mu ^2+ R^\NT \left(2 \alpha_\mu ^3+2 \alpha_\mu \right)-\alpha_\mu
   ^2 = 0,
\end{equation}
where one has to retain the only positive real solution. Then, 
\begin{equation}
    R^\mu = \sqrt{\frac{\autocvol^\NT_0}{\Xi^\mu_0}} \sqrt{1+(R^\NT)^2+2R^\NT\alpha_\mu},
\end{equation}

\begin{equation}
    \rho = \frac{R^\NT}{1+(R^\NT)^2+R^\NT\alpha_\mu}
\end{equation}
and finally
\begin{equation}
    G_0 = \sqrt{\frac{\Xi^\mu_0}{\autocvol^\NT_0}}\frac{\alpha_\mu   \sqrt{(R^\NT)^2-2 \alpha_\mu  R^\NT+1}}{(1-\alpha_\mu )
   R^\NT \left(-3 \alpha_\mu +R^\NT \left(2-\frac{1}{\left((R^\NT)^2-\alpha_\mu 
   R^\NT+1\right) \left(\sqrt{1-\frac{(R^\NT)^2}{\left((R^\NT)^2-\alpha_\mu 
   R^\NT+1\right)^2}}+1\right)}\right)+\frac{2}{R^\NT}\right)}.
\end{equation}

\section{Solution of the Markovian case}
\subsection{Construction of the Ansatz}
\label{app:construction_ansatz}

In this section we prove the results presented in Sec.~\ref{subsec:4.1}, in particular Eqs.~\eqref{eq:Markov_g_ansatz} and \eqref{eq:link_rho_Cq}.
i) First, we rewrite the property exposed in Eq.~\eqref{eq:price_corr_mark} in expectation form. ii) Then we inject in this form the quasi-camouflage property and we find a simple finite-difference equation for the propagator whose solution gives the formulas presented in Eqs.~\eqref{eq:Markov_g_ansatz} and \eqref{eq:link_rho_Cq}.

i)  
If the price ACF is exponentially decaying with the  dividends timescale, as found by means of the numerical solver, then the following relation holds: 
\begin{equation}
\label{corr p_t}
\mathbb{E}[p_{t+1}|\filt^\MM_t] = \alpha_\mu p_t.
\end{equation}

  Equation \eqref{corr p_t} gives us a relation between the excess demand ACF and the propagator. In fact using the  equation that defines the propagator model, i.e., \(p_t = \sum_{t'\leq t}G_{t-t'}q_{t'}\), it can be rewritten as:
\begin{equation}
\label{eq:preansatz}
\ G_0 \mathbb{E}[q_{t+1}|\mathcal{I}^{\MM}_t] = \alpha_\mu \sum_{t'=-\infty}^t G_{t-t'}q_{t'} - \sum_{t'=-\infty}^t G_{t+1-t'}q_{t'} .
\end{equation}
This equation is particularly interesting and  it holds regardless the structure of the NT's  trades auto-covariance.

Let us give a first example of how the above equation can be used in order to derive the result about non correlated NT's trades. 
The camouflage is exact in this case, so the excess demands are  uncorrelated, i.e., the l.h.s. of the above equation is zero, then we can see that  \(G\) decays itself exponentially with the  dividends time-scale.
This is precisely what happens if the NT are not correlated, where the propagator is  given by Eq.~\eqref{eq:prop_NT_non_corr}. 

ii) In the following we deal with the case of arbitrary Markovian NT's trades process. 
Using the expression of the general forecast matrix of a Gaussian process with zero mean, we can rewrite Eq.~\eqref{eq:preansatz}, as
\begin{equation}
\label{preansatz2}
   \left[(\tilde{\autocvol})^{-1}_0\right]^{-1} (\tilde{\autocvol})^{-1}_{t+1-t'} = \tilde{G}_{t+1-t'} -\alpha_\mu \tilde{G}_{t-t'}, \ \  \tilde{G}_{\tau} = G_\tau/ G_0.
\end{equation}

Since  we found that in generic situations an approximate camouflage relation holds, we know that the structure of the excess demand ACF matrix is given by Eq.~\eqref{eq:camouflage}. The inverse of the excess demand ACF can be computed, and it is given by:
\begin{equation}
(\tilde{\autocvol})^{-1}_0 = \frac{\omega-\sqrt{\omega^2-4}}{2\tilde{b} \alpha_\NT}, \qquad  \qquad  \omega = \frac{\tilde{b}+1+\tilde{b}\alpha_\NT^2-\alpha_\NT^2}{\tilde{b} \alpha_\NT}, \ \ 
\end{equation}
and
\begin{equation}
(\tilde{\autocvol})^{-1}_{t+1-t'} = -\frac{1}{\alpha_\NT \tilde{b}}\left( 1-(1+\tilde{b}-\alpha_\NT^2)(\tilde{\autocvol})^{-1}_0\right) \left[(\tilde{\autocvol})^{-1}_0\alpha_\NT \tilde{b} \right]^{t-t'}.
\end{equation}

Then, we can rewrite Eq.~\eqref{preansatz2} as
\begin{equation}
\label{eq:diff_eq_prop}
\tilde{G}_{t+1-t'} = \alpha_\mu \tilde{G}_{t-t'} + P \rho^{t-t'},
\end{equation}
where we defined
\begin{equation}
\label{params}
P = -\frac{1}{\alpha_\NT \tilde{b}}\left( 1-(1+\tilde{b}-\alpha_\NT^2)(\tilde{\autocvol})^{-1}_0\right) \left[(\tilde{\autocvol})^{-1}_0\right]^{-1}, \qquad \qquad \rho = (\tilde{\autocvol})^{-1}_0 \alpha_\NT \tilde{b}.
\end{equation}

The solution of Eq.~\eqref{eq:diff_eq_prop}  is  Eq.~\eqref{eq:Markov_g_ansatz} introduced in the main text. Moreover the second equation in Eqs.~\eqref{params} gives Eq.~\eqref{eq:link_rho_Cq}.

\subsection{Solving the ansatz}
\label{app:solution_ansatz}
In this appendix we present the calculations  which allowed us to obtain the results  presented in the figures of  Secs.~\ref{subsec:4.2}, \ref{subsec:4.3} and \ref{subsec:4.4}.

From the expression of the propagator given by Eq.~\eqref{eq:Markov_g_ansatz}, one is able to derive the inverse of the symmetrized propagator, which is given by
\begin{equation}
    (\tilde{\matprop}^{\text{sym}})^{-1}_{t,t'} = \Gamma_1 \gamma_1^{t-t'} +\Gamma_2 \gamma_2^{t-t'} + \delta(t-t'),   
\end{equation}
where  \(\Gamma_1\) and \(\Gamma_2\) are the solution of the following set of equations:
\begin{equation}
\begin{split}
    & \Gamma_1\frac{\alpha_\mu}{\alpha_\mu-\gamma_1} + \Gamma_2\frac{\alpha_\mu}{\alpha_\mu-\gamma_2} + 1 = 0, 
    \\ & \Gamma_1\frac{\rho}{\rho-\gamma_1} + \Gamma_2\frac{\rho}{\rho-\gamma_2} + 1 = 0, 
\end{split}
\end{equation}
whereas \(\gamma_1\) and \(\gamma_2\) are the two real positive  solution of the  equation below:
\begin{equation}
    \frac{\alpha_\mu-\alpha_\NT}{\alpha_\mu-\rho}\left(\frac{1}{1-\alpha_\mu\gamma_1}-\frac{\alpha_\mu}{\alpha_\mu-\gamma_1}\right) + \left(1- \frac{\alpha_\mu-\alpha_\NT}{\alpha_\mu-\rho}\right)\left(\frac{1}{1-\rho\gamma_1}-\frac{\rho}{\rho-\gamma_1}\right) +1= 0.
\end{equation}

With the explicit expression of \(\mathsf{G}^{sym}\) given above one is able to calculate the IT's demand Kernels given by Eqs.~\eqref{eq:IT_response}. These are given by
\begin{equation}
\begin{split}
    R_{t-t'} &= -\alpha^{t-t'}\frac{\alpha_\mu-\alpha_\NT}{\alpha_\mu-\rho}\left( \frac{\Gamma_1}{1-\gamma_1\alpha_\mu} + \frac{\Gamma_2}{1-\gamma_2\alpha_\mu}+1 \right) -\rho^{t-t'}\left(1-\frac{\alpha_\mu-\alpha_\NT}{\alpha_\mu-\rho}\right)\left( \frac{\Gamma_1}{1-\gamma_1\alpha_\mu} + \frac{\Gamma_2}{1-\gamma_2\alpha_\mu}+1 \right)
\\
    R^\NT_{t-t'} &= \delta_{t'-t} R^\NT
\\
    R^{\mu}_{t-t'} &= \delta_{t'-t} R^{\mu}
\end{split}
\end{equation}
where
\begin{equation}
\begin{split}
    R^\NT =& -\alpha_\NT \left[\frac{\alpha_\mu-\alpha_\NT}{\alpha_\mu-\rho} \left(\frac{\Gamma_1}{(1-\alpha_\mu  \gamma_1) (1-\alpha_\NT  \gamma_1)}+\frac{\Gamma_2}{(1-\alpha_\mu  \gamma_2) (1-\alpha_\NT  \gamma_2)}\right) \right.
   \\ & \left.  \ \ \  \ \ \ +\left(1-\frac{\alpha_\mu-\alpha_\NT}{\alpha_\mu-\rho}\right) \left(\frac{\Gamma_1}{(1-\alpha_\NT  \gamma_1) (1-\rho \gamma_1)}+\frac{\Gamma_2}{(1-\alpha_\NT  \gamma_2) (1-\rho \gamma_2)}\right)+1\right],
\\
    R^{\mu} =& \frac{\alpha_\mu}{G_0 (1-\alpha_\mu)}\left(\frac{\Gamma_1}{1-\gamma_1\alpha_\mu} + \frac{\Gamma_2}{1-\gamma_2\alpha_\mu}+1\right).
\end{split}
\end{equation}
Moreover, by a careful inspection of previous formulas and numerical solver results of Eq.~\eqref{eq:master_equation} in the markovian case, one realize that the following property holds:
\begin{equation}
    R^{\mu} = \sqrt{\frac{\autocvol_0^\NT}{\Xi^\mu_0}}\sqrt{({R^\NT})^2+2 \alpha_\NT  R^\NT+1}.
\end{equation}
From the equation above one is able to deduce the expression of \(G_0\), by inverting the previous equation for \(R^{\mu}\).

Finally, imposing the break even condition per trade of the MM given by Eq.~\eqref{eq:avg_break_even}, one is able to derive the following identity:
\begin{equation}
\label{eq:Omega_0}
    \Omega_0 = \Xi^\mu_0 (R^\mu)^2\frac{\alpha_\mu   \rho }{\gamma_1\gamma_2 } \left(\tilde{b}+\frac{\alpha_\mu-\alpha_\NT}{\alpha_\mu-\rho}\frac{1}{1-\alpha_\mu  \alpha_\NT } +\left(1-\frac{\alpha_\mu-\alpha_\NT}{\alpha_\mu-\rho}\right)\frac{1}{1-\alpha_\NT  \rho }\right).
\end{equation}
In order to close the ansatz on itself we have to compute the total order flow ACF. To do this, we need to calculate the first row of the inverse \((\id-\mathsf{R}\matL)^{-1}\) which appear in Eq.~\eqref{eq:excess_demand_dynamics}.
This is given by
\begin{equation}
    \{(\id-\matdmd \matL)^{-1}\}_{t-t'} = \frac{\{(\mathsf{G}^{sym})^{-1}\}_{t,t'}}{\{(\mathsf{G}^{sym})^{-1}\}_{t,t}} = \frac{\alpha_\mu \rho}{\gamma_1\gamma_2}\{\tilde{\mathsf{G}}^{sym}\}_{t-t'}.
\end{equation}

The explicit expression of the excess demand at time $t$ 
is given by
\begin{equation}
\begin{split}
    q_t = \frac{\alpha_\mu \rho}{\gamma_1\gamma_2} &\left\{  \left[  q^{\NT}_t +\sum_{t'=-\infty}^t\left( \Gamma_1 \gamma_1^{t-t'} + \Gamma_2\gamma_2^{t-t'}\right)q^{\NT}_{t'}\right] \right.
    \\ & + 
    R_\NT\left[q^{\NT}_{t-1} +\sum_{t'=-\infty}^{t-1}\left( \Gamma_1 \gamma_1^{t-t'-1} + \Gamma_2\gamma_2^{t-t'-1}\right)q^{\NT}_{t'}\right]
    \\&  + \left.
    R_{\mu}\left[\mu_{t-1} +\sum_{t'=-\infty}^{t-1}\left( \Gamma_1 \gamma_1^{t-t'-1} + \Gamma_2\gamma_2^{t-t'-1}\right)\mu_{t'}\right] \right\}. 
\end{split}
\end{equation}

With this equation one is able to compute explicitly the excess demand ACF. In particular, by comparing the lag-0 term of it with the functional form given in Eq.~\eqref{eq:camouflage} and using Eqs.~\eqref{eq:link_rho_Cq} and \eqref{eq:Omega_0} one is able to compute an implicit very complicated equation for $\rho$, fixing completely the ansatz given by Eq.~\eqref{eq:Markov_g_ansatz}.

The figures presented  in Sec.~\ref{sec:Markov} have been obtained by fitting the result of the numerical solver with Eq.~\eqref{eq:Markov_g_ansatz}, obtaining numerical values for $\rho$ which have been cross-validated using the aformentioned analytical implicit equation for $\rho$, and then using the equations exposed in this section to compute the other quantities of interest.

\end{document}